\documentclass[10pt,twocolumn,english,aps]{revtex4-1}
\usepackage{ae,aecompl}

\usepackage[T1]{fontenc}
\usepackage[latin9]{inputenc}
\usepackage{geometry}
\usepackage{amsmath}
\usepackage{amssymb}
\usepackage{graphicx}

\makeatletter
\usepackage{braket}

\makeatother

\usepackage{babel}
\begin{document}
\global\long\def\ket#1{\left|#1\right\rangle }
\global\long\def\bra#1{\left\langle #1\right|}
\global\long\def\braket#1#2{\langle#1|#2\rangle}

\title{Gauge covariances and nonlinear optical responses}

\author{G. B. Ventura}
\email{corresponding author: gbventura@fc.up.pt}

\author{D. J. Passos}

\affiliation{Centro de F\'isica das Universidades do Minho e Porto}

\affiliation{Departamento de F\'isica e Astronomia, Faculdade de Ci\^encias,
Universidade do Porto, 4169-007 Porto, Portugal}

\author{J. M. Viana Parente Lopes}

\affiliation{Centro de F\'{i}sica das Universidades do Minho e Porto}

\affiliation{Departamento de Engenharia Física, Faculdade de Engenharia}

\affiliation{Departamento de F\'isica e Astronomia, Faculdade de Ci\^encias,
Universidade do Porto, 4169-007 Porto, Portugal}

\author{J. M. B. Lopes dos Santos}

\affiliation{Centro de F\'isica das Universidades do Minho e Porto}

\affiliation{Departamento de F\'isica e Astronomia, Faculdade de Ci\^encias,
Universidade do Porto, 4169-007 Porto, Portugal}

\author{N. M. R. Peres}

\affiliation{Centro de F\'isica das Universidades do Minho e Porto}

\affiliation{Departamento de F\'isica, Universidade do Minho, P-4710-057, Braga,
Portugal}
\begin{abstract}
The formalism of the reduced density matrix is pursued in both length
and velocity gauges of the perturbation to the crystal Hamiltonian.
The covariant derivative is introduced as a convenient representation
of the position operator. This allow us to write compact expressions
for the reduced density matrix in any order of the perturbation which
simplifies the calculations of nonlinear optical responses; as an
example, we compute the first and third order contributions of the
monolayer graphene. Expressions obtained in both gauges share the
same formal structure, allowing a comparison of the effects of truncation
to a finite set of bands. This truncation breaks the equivalence between
the two approaches: its proper implementation can be done directly
in the expressions derived in the length gauge, but require a revision
of the equations of motion of the reduced density matrix in the velocity
gauge.
\end{abstract}
\maketitle

\section{Introduction}

The calculation of nonlinear optical (NLO) coefficients in crystals
has seen a renewed impetus, spurred by the strong nonlinear properties
of layered materials like graphene \cite{0953-8984-20-38-384204,Bonaccorso2010,Glazov2014b,Cheng2014,PhysRevB.90.125425,PhysRevB.91.235320,PhysRevB.92.235432,Mikhailov2016}.

Perturbative calculations of NLO coefficients in bulk semiconductors,
with a full quantum treatment of matter, date back to early nineties
of the previous century, and have not been entirely trouble free \cite{Moss1990,Sipe1993}.
In the long wavelength limit\textemdash in which the spatial dependence
of the radiation electric field is neglected\textemdash , there are
two representations of the radiation field: by a time dependent vector
potential $\mathbf{A}(t)$, with the electric field given by $\mathbf{E}(t)=-\partial\mathbf{A}/\partial t$;
by the electric dipole scalar potential, $V(\mathbf{r})=e\mathbf{E}(t)\cdot\mathbf{r}$.
The advantage of the first method, known as the velocity gauge, is
that the perturbation introduces no extra spatial dependence to the
crystal Hamiltonian, thus preserving the crystal's translational symmetry.
This leads to a decoupling of the system's response in momentum space:
it becomes a sum of independent contributions of each $\mathbf{k}$
value in the Brillouin zone. Early attempts to calculate NLO coefficients
using this approach were, however, plagued by unphysical contributions,
diverging at low frequencies \cite{Moss1990}. Several authors addressed
this issue by separating the treatment of inter and intra band contributions,
using time-dependent basis sets \cite{Genkin1968b,Sipe1993}. Later
Aversa and Sipe \cite{Aversa1995} revisited the problem, emphasizing
the gauge freedom that allows you to choose either form of the coupling
to the radiation field. They recognized that the unphysical divergences
mentioned above actually have coefficients that are exactly zero,
expressing sum rules that they derived explicitly in first order response,
and claimed to hold in all orders. Because these sums rules are easily
violated in approximations, they end up advocating using the scalar
potential method, also referred to as the length gauge, in actual
calculations.

Similar problems were found in earlier calculations of NLO response
of atoms \cite{PhysRevA.36.2763}. As far back as 1951, ref.\,\cite{PhysRev.85.259},
W. Lamb Jr. recommended as more convenient the scalar potential gauge
in perturbative calculations of the Hydrogen atom fine structure,
and, for a while, the view that the two choices of gauge lead to different
results was widely held \cite{PhysRevA.36.2763}. 

The obvious advantage of the scalar potential gauge is that it is
written in terms of a gauge invariant entity, the electric field,
even though it expresses a specific choice of gauge for the electromagnetic
field. But because the scalar potential contains the position variable,
$\mathbf{r}$, the perturbation is no longer diagonal in Bloch momentum
space, and couples different $\mathbf{k}$ values. Furthermore, the
position operator is highly singular in momentum space, and its matrix
elements can only be properly defined in the infinite crystal limit.

This choice of representation was used on the recent reduced density
matrix (RDM) calculations of the nonlinear optical response of graphene
\cite{Cheng2014,PhysRevB.91.235320,Mikhailov2016}. In ref.\,\cite{Cheng2014},
the derivative term in the RDM equations of motion was removed by
means of a \textbf{$\mathbf{k}$}-space translation thus decoupling
them in crystal momentum space; this is not without cost, as the system's
response is now expressed in terms of both the \textbf{$\mathbf{A}$}
and the \textbf{$\mathbf{E}$ }fields. In their subsequent work \cite{PhysRevB.91.235320},
the authors retained this derivative term as well as introduced relaxation
terms to the RDM equations of motion. A different approach was proposed
by Mikhailov \cite{Mikhailov2016}, who avoided the problem of the
singular intra band term of the position operator by using a finite
wavelength perturbation that satisfies periodic boundary conditions,
\begin{equation}
V(\mathbf{r})=-e\left(V_{\mathbf{q}}e^{i\mathbf{q}\cdot\mathbf{r}}+V_{\mathbf{q}}^{*}e^{-i\mathbf{q}\cdot\mathbf{r}}\right),
\end{equation}
and taking the limit $\mathbf{q}\to0$ at the end of the calculation.

In addition to the freedom in expressing the external electric field,
we consider the freedom of choice of the phase of the Bloch functions,
in each point in the Brillouin zone, 
\[
\psi_{\mathbf{k}s}\rightarrow e^{i\theta_{s}(\mathbf{k})}\,\psi_{\mathbf{k}s},
\]
since any expectation value is necessarily independent from the choice
of $\theta_{s}(\mathbf{k})$. This sets the transformation law of
any observable's matrix elements to
\[
\mathcal{A}_{\mathbf{k}\mathbf{k}'ss'}\rightarrow e^{-i(\theta_{s}(\mathbf{k})-\theta_{s'}(\mathbf{k}'))}\,\mathcal{A}_{\mathbf{k}\mathbf{k}'ss'}.
\]
Striving to make this property explicit leads to the concept of the
covariant derivative, which will be used to derive a considerably
simpler form of the RDM formalism in the length gauge.

This paper is organized as follows. In Section\,II, we present an
overview of some concepts on gauge invariance and the key ideas regarding
the crystal Hamiltonian. We then introduce the covariant derivative
in momentum space, which captures the phase freedom mentioned in the
previous paragraph. Section\,III is dedicated to the reduced density
matrix. We shall derive the RDM equations of motions and, more importantly,
the relation between these two objects. The rest of the section is
dedicated to the equivalence between observables in the two formalisms.
In Section\,IV, we write the solutions to the RDM equations of motion.
As a proof of concept, Section\,V is dedicated to the study of graphene's
current response by means of the scalar potential formalism \cite{Cheng2014,PhysRevB.91.235320,Mikhailov2016}.
This is followed by Section\,VI, where we discuss the breakdown of
the scalar potential/vector potential equivalence, upon truncation
of the expressions for the current for a finite set of bands \cite{boyd2004}.
The last section is dedicated to a summary of our results.

\section{A GENERAL DESCRIPTION }

The two possible representations of the uniform electric field entail
two different, but equivalent, ways to write the many-body Hamiltonian.
One can either add the dipole interaction to the single particle Hamiltonian
of the unperturbed system, $\mathcal{H}_{0}$, \footnote{The charge $q$ has already been replaced by the charge of an electron,
$q=-e$.} 
\begin{equation}
H_{E}(t)=\int d^{d}\mathbf{r}\,\Psi^{\dagger}(\mathbf{r})\left[\mathcal{H}_{0}\left(\mathbf{r},\frac{\nabla}{i}\right)+e\mathbf{E}(t)\cdot\mathbf{r}\right]\Psi(\mathbf{r}),\label{eq:HAM1}
\end{equation}
or use the minimum coupling procedure in $\mathcal{H}_{0}$,
\begin{equation}
H_{A}(t)=\int d^{d}\mathbf{r}\,\Psi^{\dagger}(\mathbf{r})\left[\mathcal{H}_{0}\left(\mathbf{r},\frac{\nabla}{i}+\frac{e}{\hbar}\mathbf{A}(t)\right)\right]\Psi(\mathbf{r}).\label{eq:HAM2}
\end{equation}
The electron field, $\Psi(\mathbf{r})$, and its Hermitian conjugate,
$\Psi^{\dagger}(\mathbf{r})$, satisfy the usual anti-commutation
relations.

The many-body state vector in the vector potential approach, $\left|\psi(t)\right\rangle $,
evolves in time according to the Hamiltonian $H_{A}(t)$
\begin{align}
i\hbar\frac{\partial\left|\psi(t)\right\rangle }{\partial t} & =H_{A}(t)\left|\psi(t)\right\rangle .\label{eq:HAM3}
\end{align}
A second state vector $\left|\bar{\psi}(t)\right\rangle $, obtained
via a time-dependent unitary transformation of $\left|\psi(t)\right\rangle $,
\begin{align}
\left|\bar{\psi}(t)\right\rangle  & =\mathcal{U}(t)\left|\psi(t)\right\rangle ,\label{eq:UNITARY}
\end{align}
has an equation of motion 
\[
i\hbar\frac{\partial\left|\bar{\psi}(t)\right\rangle }{\partial t}=\left[\mathcal{U}(t)H_{A}(t)\mathcal{U}^{\dagger}(t)+i\hbar\frac{d\mathcal{U}(t)}{dt}\mathcal{U}^{\dagger}(t)\right]\left|\bar{\psi}(t)\right\rangle .
\]
If the unitary transformation is chosen as 
\begin{equation}
\mathcal{U}(t)=\exp\left[i\frac{e}{\hbar}\int d^{d}\mathbf{r}\,\mathbf{A}(t)\cdot\mathbf{r}\,\rho(\mathbf{r})\right],\label{eq:HAM4}
\end{equation}
where $\rho(\mathbf{r}):=\Psi^{\dagger}(\mathbf{r})\Psi(\mathbf{r})$
is the density operator, it is straightforward to show that 
\begin{equation}
\mathcal{U}(t)\,H_{A}(t)\,\mathcal{U}^{\dagger}(t)+i\hbar\frac{d\mathcal{U}(t)}{dt}\,\mathcal{U}^{\dagger}(t)=H_{E}(t),\label{eq:HAM_U}
\end{equation}
implying that $\left|\bar{\psi}(t)\right\rangle $ is the state vector
in scalar potential gauge.

Observables in the two gauges are also related by a unitary transformation,
$O_{E}:=\mathcal{U}(t)\,O_{A}(t)\,\mathcal{U}^{\dagger}(t)$. The
exception is the Hamiltonian; because the unitary transformation,
$\mathcal{U}(t)$, is time dependent, the time evolution operator
in the length gauge, $H_{E}(t)$, is not simply $\mathcal{U}(t)\,H_{A}(t)\,\mathcal{U}^{\dagger}(t)$,
but has an additional term involving the time derivative of  $\mathcal{U}(t)$,
Eq.\,(\ref{eq:HAM_U}). The existence of this transformation between
the two descriptions of the radiation field establishes their complete
equivalence \cite{PhysRevA.36.2763,Aversa1995}. 

Next we recall some important results of electron eigenstates in an
unperturbed crystal. The single particle Schr\"{o}dinger equation
is \cite{Ashcroft1976} 
\begin{equation}
\mathcal{H}\psi_{\mathbf{k}s}(\mathbf{r})=\epsilon_{\mathbf{k}s}\psi_{\mathbf{k}s}(\mathbf{r}),\label{eq:CRYSTALSCHRO}
\end{equation}
with 
\begin{equation}
\mathcal{H}=\frac{\hbar^{2}}{2m}\left(\frac{\nabla}{i}\right)^{2}+V(\mathbf{r}),\label{eq:schrodinger_ham}
\end{equation}
and $V(\mathbf{r})=V(\mathbf{r}+\mathbf{R})$, for $\mathbf{R}$ any
Bravais lattice vector. According to Bloch's theorem, the eigenfunctions
have the form of a plane wave times a periodic function, 
\begin{equation}
\psi_{\mathbf{k}s}(\mathbf{r})=e^{i\mathbf{k}\cdot\mathbf{r}}u_{\mathbf{k}s}(\mathbf{r}),\label{eq:CH1}
\end{equation}
allowing the eigenvalue problem to be expressed in terms of the $\mathbf{k}$-dependent
Hamiltonian, $\mathcal{H}(\mathbf{k}):=e^{-i\mathbf{k}\cdot\mathbf{r}}\,\mathcal{H}\,e^{i\mathbf{k}\cdot\mathbf{r}}$,
\begin{eqnarray}
\mathcal{H}(\mathbf{k})\,u_{\mathbf{k}s}(\mathbf{r}) & = & \left[\frac{\hbar^{2}}{2m}\left(\frac{\nabla}{i}+\mathbf{k}\right)^{2}+V(\mathbf{r})\right]u_{\mathbf{k}s}(\mathbf{r})\nonumber \\
 & = & \epsilon_{\mathbf{k}s}u_{\mathbf{k}s}(\mathbf{r}).\label{eq:CH3}
\end{eqnarray}
The function $u_{\mathbf{k}s}(\mathbf{r})$ is a periodic function
in the real space unit cell,
\begin{equation}
u_{\mathbf{k}s}(\mathbf{r})=u_{\mathbf{k}s}(\mathbf{r}+\mathbf{R}).\label{eq:CH2}
\end{equation}
Each $\mathbf{k}$-point in the First Brillouin Zone (FBZ) defines
an Hamiltonian operator, $\mathcal{H}(\mathbf{k})$, that acts on
functions whose domain is the real space unit cell, and that satisfy
the boundary condition of Eq.\,(\ref{eq:CH2}). The eigenfunctions
of $\mathcal{H}_{\mathbf{k}}$, $\left\{ \ket{u_{\mathbf{k}s}},\,s=0,1,\dots\right\} $
are a basis of such functions. We can assume this basis to be orthonormal
with an inner product defined as the integral over the real space
unit cell (volume $v_{c}$),
\begin{equation}
\braket{u_{\mathbf{k}s}}{u_{\mathbf{k}s'}}=\frac{1}{v_{c}}\int_{uc}d^{d}\mathbf{r}\,u_{\mathbf{k}s}^{*}\left(\mathbf{r}\right)u_{\mathbf{k}s'}\left(\mathbf{r}\right)=\delta_{ss'}.\label{eq:CH4}
\end{equation}
Different values of $\mathbf{k}$ have different basis, for they are
eigenfunctions of different Hamiltonians. Here, the Bloch wave vector
is a continuous parameter, even in the finite volume crystal, as the
eigenvalues in Eq\@.\,(\ref{eq:CH3}) are well defined for every
$\mathbf{k}$ in the FBZ. The $\mathbf{k}$ value selection by periodic
boundary conditions only involves the plane wave factor of the Bloch
function, and has no bearing on the periodic part, $u_{\mathbf{k}s}\left(\mathbf{r}\right)$
\cite{Ashcroft1976}. As such, derivatives with respect to $\mathbf{k}$
of $u_{\mathbf{k}s}\left(\mathbf{r}\right)$ are always well defined
whereas derivatives of plane wave factors require the infinite volume
limit. We shall work in this limit from the start, due to the difficulties
of properly defining the position operator, $\mathbf{r}$, in a finite
system with periodic boundary conditions. 

In the $\Omega\to\infty$ limit, ($\Omega$, the volume of the crystal)
the momentum sums are replaced by $d$-dimensional integrals over
the FBZ. The many-body crystalline Hamiltonian then reads as 
\begin{eqnarray}
H_{0} & = & \sum_{s}\int\frac{d^{d}\mathbf{k}}{(2\pi)^{d}}\epsilon_{\mathbf{k}s}c_{\mathbf{k}s}^{\dagger}c_{\mathbf{k}s},\label{eq:CH5}
\end{eqnarray}
for $c_{\mathbf{k}s}^{\dagger}$, $c_{\mathbf{k}s}$ the creation
and destruction operators of Bloch states, 
\begin{align}
c_{\mathbf{k}s}^{\dagger} & =\int d^{d}\mathbf{r}\,\psi_{\mathbf{k}s}(\mathbf{r})\,\Psi^{\dagger}(\mathbf{r}).
\end{align}
The Bloch state orthogonality relation, the anti-commutation relations,
and the lattice sum rule are suitably modified,
\begin{align}
\braket{\psi_{\mathbf{k}s}}{\psi_{\mathbf{k}'s'}} & =(2\pi)^{d}\delta_{ss'}\delta(\mathbf{k}-\mathbf{k}'),\label{eq:CH6}
\end{align}
\begin{equation}
\bigl\{ c_{\mathbf{k}s},c_{\mathbf{k'}s'}^{\dagger}\bigr\}=(2\pi)^{d}\delta_{ss'}\delta(\mathbf{k}-\mathbf{k}'),
\end{equation}
\begin{equation}
\sum_{\mathbf{R}}e^{i(\mathbf{k}-\mathbf{k}')\cdot\mathbf{R}}=\frac{(2\pi)^{3}}{v_{C}}\delta(\mathbf{k}-\mathbf{k}'),
\end{equation}
so that the operator $c_{\mathbf{k}s}^{\dagger}c_{\mathbf{k}s}$ is
a density in momentum space and not a dimensionless number operator
as in the finite volume case. 

In the scalar potential approach, the perturbation is written in terms
of the position operator, $\mathbf{r}$. Its matrix elements are ill-defined
in the finite volume system, but can be computed for $\Omega\rightarrow\infty$.
In that limit they read \cite{Blount1962.},
\begin{eqnarray}
\mathbf{r}{}_{\mathbf{k}\mathbf{k}',ss'} & = & \delta_{ss'}\,(2\pi)^{d}(-i)\nabla_{\mathbf{k}'}\delta(\mathbf{k}'-\mathbf{k})\nonumber \\
 &  & +(2\pi)^{d}\delta(\mathbf{k}'-\mathbf{k})\,\boldsymbol{\xi}_{\mathbf{k}'ss'},\label{eq:CH8}
\end{eqnarray}
where the Berry connection, $\boldsymbol{\xi}_{\mathbf{k}ss'}$, is
defined as a scalar product in the real space unit cell, independent
of the crystal's volume, 
\begin{eqnarray}
\boldsymbol{\xi}_{\mathbf{k}ss'} & := & i\braket{u_{\mathbf{k}s}}{\nabla_{\mathbf{k}}u_{\mathbf{k}s'}}\label{eq:CH9}\\
 & = & \frac{i}{v_{C}}\int_{uc}d^{d}\mathbf{r}\,u_{\mathbf{k}s}^{*}(\mathbf{r})\nabla_{\mathbf{k}}u_{\mathbf{k}s'}(\mathbf{r}).\label{eq:CH10}
\end{eqnarray}
The somewhat awkward looking expression of Eq.\,(\ref{eq:CH8}),
can be cast in a more transparent form if we bear in mind that, for
a continuous non-normalizable basis, the matrix elements of an operator
are a kernel of an integral transform. In the Bloch representation,
a general single particle state is represented as 
\[
\Psi(\mathbf{r})=\sum_{s}\int\frac{d^{d}\mathbf{k}}{(2\pi)^{d}}\Phi_{s}\left(\mathbf{k}\right)\psi_{\mathbf{k}s}(\mathbf{r}),
\]
for $\Phi_{s}\left(\mathbf{k}\right)=\braket{\psi_{\mathbf{k}s}}{\Psi}$.
The wave function for the state $\mathbf{r}\ket{\Psi}$ is 
\begin{align*}
\bra{\psi_{\mathbf{k}s}}\mathbf{r}\ket{\Psi} & =\sum_{s'}\int\frac{d^{d}\mathbf{k'}}{(2\pi)^{d}}\mathbf{r}{}_{\mathbf{k}\mathbf{k}',ss'}\Phi_{s'}\left(\mathbf{k'}\right)\\
=i\sum_{s'} & \left(\delta_{ss'}\nabla_{\mathbf{k}}-i\boldsymbol{\xi}_{\mathbf{k}ss'}\right)\Phi_{s'}\left(\mathbf{k}\right),
\end{align*}
where, to reach the final expression, we have integrated by parts.
The integration is over the FBZ, and, given the periodicity of any
function of $\mathbf{k}$, $\phi(\mathbf{k})=\phi(\mathbf{k}+\mathbf{G})$,
there are no surface terms. This prompts us do define the covariant
derivative operator \cite{PhysRevB.91.235320}, 
\begin{equation}
\mathbf{D}_{\mathbf{k}ss'}:=\delta_{ss'}\nabla_{\mathbf{k}}-i\boldsymbol{\xi}_{\mathbf{k}ss'},\label{eq:cov_derivative}
\end{equation}
such that 
\begin{equation}
\bra{\psi_{\mathbf{k}s}}\mathbf{r}\ket{\Psi}=\sum_{s'}i\mathbf{D}_{\mathbf{k}ss'}\Phi_{s'}\left(\mathbf{k}\right),\label{eq:postion_cov_derivatiove}
\end{equation}
\emph{i.e., }the position operator is $i\mathbf{D}_{\mathbf{k}ss'}$
in the Bloch representation \cite{LandauStat2}. The designation of
\emph{covariant }refers to its behavior under a local gauge transformation
in momentum space, 
\begin{equation}
u_{\mathbf{k}s}\rightarrow e^{i\theta_{s}(\mathbf{k})}\,u_{\mathbf{k}s},\label{eq:gauge_k_space}
\end{equation}
for which 
\begin{align*}
\mathbf{D}_{\mathbf{k}ss'} & \to\tilde{\mathbf{D}}_{\mathbf{k}ss'}:=e^{-i\theta_{s}(\mathbf{k})}\mathbf{D}_{\mathbf{k}ss'}e^{i\theta_{s'}(\mathbf{k})}.\\
 & =e^{-i(\theta_{s}(\mathbf{k})-\theta_{s'}(\mathbf{k}))}\mathbf{D}_{\mathbf{k}ss'}.
\end{align*}
The gradient term of the phase $\theta_{s'}(\mathbf{k})$ is canceled
by the transformation of the Berry connection; this is in complete
parallel to the definition of covariant derivative in gauge theories
in real space. 

In the vector potential approach, the perturbation is written in terms
of the velocity matrix elements, $\mathbf{v}{}_{\mathbf{k}\mathbf{k}'ss'}:=\mathbf{p}{}_{\mathbf{k}\mathbf{k}'ss'}/m_{e}$,
which are diagonal in\textbf{ }$\mathbf{k}$-space, and expressible
as matrix elements in the basis of periodic functions,
\begin{align}
\mathbf{v}_{\mathbf{k}\mathbf{k}'ss'} & =(2\pi)^{d}\delta(\mathbf{k}-\mathbf{k}')\,\mathbf{v}_{\mathbf{k}ss'},\label{eq:CH12}\\
\mathbf{v}_{\mathbf{k}ss'} & =\frac{\hbar}{m_{e}}\bra{u_{\mathbf{k}s}}(-i\nabla+\mathbf{k})\ket{u_{\mathbf{k}s'}},\label{eq:CH13}
\end{align}
Since the velocity operator is quite generally $(i\hbar)^{-1}\left[\mathbf{r},H\right]$,
it is not surprising to find that 
\begin{align}
\mathbf{v}_{\mathbf{k}ss'} & =\frac{1}{\hbar}\left[\mathbf{D}_{\mathbf{k}},\mathcal{H}(\mathbf{k})\right]_{ss'},\label{eq:VELCOM}\\
 & =\frac{1}{\hbar}\left[\delta_{ss'}\nabla_{\mathbf{k}}\epsilon_{\mathbf{k}s}-i(\epsilon_{\mathbf{k}s'}-\epsilon_{\mathbf{k}s})\boldsymbol{\xi}_{\mathbf{k}ss'}\right],\label{eq:VELCOM1}
\end{align}
This turns out to be the expression of a more general result, which
will prove useful later and which we now discuss.

Any matrix in the space generated by the basis $\left\{ \ket{u_{\mathbf{k}s}},\,s=1,2,\dots\right\} $
, parametrized by $\mathbf{k}$, defines an operator in band space
by 

\begin{align}
\mathcal{\mathcal{O}}(\mathbf{k}) & =\sum_{ss'}\ket{u_{\mathbf{k}s}}\mathcal{\mathcal{O}}_{\mathbf{k}ss'}\bra{u_{\mathbf{k}s'}},\label{eq:BISMAT-1}\\
\mathcal{\mathcal{O}}_{\mathbf{k}ss'} & =\left\langle u_{\mathbf{k}s}\right|\mathcal{\mathcal{O}}(\mathbf{k})\left|u_{\mathbf{k}s'}\right\rangle .\label{eq:BISMATELE-1}
\end{align}
Two examples of this are the Hamiltonian, $\mathcal{H}(\mathbf{k}):=\epsilon_{\mathbf{k}s}\delta_{ss'}$
and the velocity operator. Consider the matrix elements of the \textbf{$\mathbf{k}$}-derivative
of $\mathcal{\mathcal{O}}(\mathbf{k})$, 
\begin{eqnarray}
\left[\nabla_{\mathbf{k}}\mathcal{\mathcal{O}}(\mathbf{k})\right]_{ss'} & := & \left\langle u_{\mathbf{k}s}\right|\nabla_{\mathbf{k}}\mathcal{\mathcal{O}}(\mathbf{k})\left|u_{\mathbf{k}s'}\right\rangle ,\nonumber \\
 & = & \nabla_{\mathbf{k}}\left\langle u_{\mathbf{k}s}\right|\mathcal{\mathcal{O}}(\mathbf{k})\left|u_{\mathbf{k}s'}\right\rangle \nonumber \\
 & - & \left\langle \nabla_{\mathbf{k}}u_{\mathbf{k}s}\right|\mathcal{\mathcal{O}}(\mathbf{k})\left|u_{\mathbf{k}s'}\right\rangle \nonumber \\
 & - & \left\langle u_{\mathbf{k}s}\right|\mathcal{\mathcal{O}}(\mathbf{k})\left|\nabla_{\mathbf{k}}u_{\mathbf{k}s'}\right\rangle .\label{eq:BISMAT2-1}
\end{eqnarray}
The first term on the right hand side is simply the \textbf{$\mathbf{k}$}-gradient
of the matrix element, $\nabla_{\mathbf{k}}\mathcal{\mathcal{O}}_{\mathbf{k}ss'}$.
The other two can be expressed in terms of the Berry connection, by
application of the completeness relations $\sum_{r}$$\ket{u_{\mathbf{k}r}}\bra{u_{\mathbf{k}r}}=\hat{1}$,
\begin{eqnarray}
\left\langle \nabla_{\mathbf{k}}u_{\mathbf{k}s}\right|\mathcal{\mathcal{O}}(\mathbf{k})\left|u_{\mathbf{k}s'}\right\rangle  & = & i\sum_{r}\boldsymbol{\xi}_{\mathbf{k}sr}\mathcal{\mathcal{O}}_{\mathbf{k}rs'},\label{eq:BISMAT3-1}\\
\left\langle u_{\mathbf{k}s}\right|\mathcal{\mathcal{O}}(\mathbf{k})\left|\nabla_{\mathbf{k}}u_{\mathbf{k}s'}\right\rangle  & = & -i\sum_{r}\mathcal{\mathcal{O}}_{\mathbf{k}sr}\boldsymbol{\xi}_{\mathbf{k}rs'}.\label{eq:BISMAT4-1}
\end{eqnarray}
The matrix element, Eq.\,(\ref{eq:BISMAT2-1}), then reads as a commutator
with the covariant derivative, 
\begin{eqnarray}
\left[\nabla_{\mathbf{k}}\mathcal{\mathcal{O}}(\mathbf{k})\right]_{ss'} & = & \nabla_{\mathbf{k}}\mathcal{\mathcal{O}}_{\mathbf{k}ss'}-i\left[\boldsymbol{\xi}_{\mathbf{k}},\mathcal{\mathcal{O}}(\mathbf{k})\right]_{ss'},\label{eq:COM1-1}\\
 & = & \bigl[\mathbf{D}_{\mathbf{k}},\mathcal{\mathcal{O}}(\mathbf{k})\bigr]_{ss'}.\label{eq:COM2-1}
\end{eqnarray}
which can be alternatively represented in operator form,
\begin{align}
\nabla_{\mathbf{k}}\mathcal{\mathcal{O}}(\mathbf{k}) & =\bigl[\mathbf{D}_{\mathbf{k}},\mathcal{\mathcal{O}}(\mathbf{k})\bigr].\label{eq:COM3-1}
\end{align}
The relation between observables in the velocity and length gauges
can be cast in this language. First, we note that
\begin{align}
\bra{\psi_{\mathbf{k}s}}\mathcal{O}^{A}\left(\frac{\nabla}{i}\right)\ket{\psi_{\mathbf{k}'s'}} & =\bra{\psi_{\mathbf{k}s}}\mathcal{O}^{E}\left(\frac{\nabla}{i}+\frac{e}{\hbar}\mathbf{A}(t)\right)\ket{\psi_{\mathbf{k}'s'}}.
\end{align}
Using the form of the Bloch functions, Eq.\,(\ref{eq:CH1}), this
integral over all space can reduced to one over a unit cell, summed
over all of them, giving, 
\begin{align}
(2\pi)^{3}\delta(\mathbf{k}-\mathbf{k}')\bra{u_{\mathbf{k}s}}\mathcal{O}^{A}\left(\mathbf{k}\right)\ket{u_{\mathbf{k}s'}}\\
=(2\pi)^{3}\delta(\mathbf{k}-\mathbf{k}')\bra{u_{\mathbf{k}s}}\mathcal{O}^{E}\left(\mathbf{k}+\frac{e}{\hbar}\mathbf{A}(t)\right)\ket{u_{\mathbf{k}s'}}
\end{align}
or, 
\begin{align}
\mathcal{O}^{A}\left(\mathbf{k}\right) & =\mathcal{O}^{E}\left(\mathbf{k}+\frac{e}{\hbar}\mathbf{A}(t)\right).\label{eq:EQUIBANDS}
\end{align}
The relationship between operators in the two descriptions can also
be expressed in terms of these objects defined in the same \textbf{k}-point.
To do so, we expand the RHS of that equation in powers of $\mathbf{A}(t)$.
It follows from Eq.\,(\ref{eq:COM3-1}), that \begin{widetext}
\begin{eqnarray}
\mathcal{O}^{A}\left(\mathbf{k},t\right)=\mathcal{O}^{E}\left(\mathbf{k}+\frac{e}{\hbar}\mathbf{A}(t),t\right) & = & \sum_{n=0}^{+\infty}\frac{1}{n!}\left(\frac{e}{\hbar}\right)^{n}A^{\alpha_{1}}(t)\dots A^{\alpha_{n}}(t)\bigl[D_{\mathbf{k}}^{\alpha_{1}},\bigl[...,\bigl[D_{\mathbf{k}}^{\alpha_{n}},\mathcal{O}^{E}\left(\mathbf{k},t\right)\bigr]...\bigr]\bigr].\label{eq:EQUI13}
\end{eqnarray}
\end{widetext} where a sum over the repeated cartesian indexes $\alpha_{j}$
is left implied. To conclude this brief account of the use of the
covariant derivative, we point out that the canonical commutation
relation, 
\begin{equation}
\bigl[\hat{r}^{\alpha},\hat{p}^{\beta}\bigr]=i\hbar\delta^{\alpha\beta}\hat{1},\label{eq:CANONICAL-1}
\end{equation}
is expressed in the Bloch basis as 
\begin{eqnarray}
\bigl[D_{\mathbf{k}}^{\alpha},\mathcal{V}{}^{\beta}(\mathbf{k})\bigr]_{ss'} & = & \frac{\hbar}{m_{e}}\delta^{\alpha\beta}\delta_{ss'},\label{eq:BISCANON-1}
\end{eqnarray}
since $\hat{r}^{\alpha}=iD^{\alpha}$ and $\hat{p}^{\alpha}=m_{e}\mathcal{V}^{\alpha}$.
Eq.\,(\ref{eq:BISCANON-1}) can also be explicitly derived from the
form of the $\mathbf{D}_{\mathbf{k}}$ and $\mathbf{v}_{\mathbf{k}}$
matrices. 

We can now write the general many-body Hamiltonians, (\ref{eq:HAM1})
and (\ref{eq:HAM2}), in the Bloch description, 
\begin{eqnarray}
H_{E}(t) & = & \sum_{ss'}\int\frac{d^{d}\mathbf{k}}{(2\pi)^{d}}c_{\mathbf{k}s}^{\dagger}\left[\delta_{ss'}\epsilon_{\mathbf{k}s}+ie\mathbf{E}(t)\cdot\mathbf{D}_{\mathbf{k}ss'}\right]c_{\mathbf{k}s'},\nonumber \\
\label{eq:HAMB1}\\
H_{A}(t) & = & \sum_{ss'}\int\frac{d^{d}\mathbf{k}}{(2\pi)^{d}}c_{\mathbf{k}s}^{\dagger}\left[\delta_{ss'}\left(\epsilon_{\mathbf{k}s}+\frac{e^{2}A^{2}(t)}{2m_{e}}\right)\right.\nonumber \\
 &  & \left.+e\mathbf{A}(t)\cdot\mathbf{v}_{\mathbf{k}ss'}\right]c_{\mathbf{k}s'}.\label{eq:HAMB2}
\end{eqnarray}
as well as their respective current operators, 
\begin{align}
\mathbf{J}_{E}(t) & =-e\sum_{ss'}\int\frac{d^{d}\mathbf{k}}{(2\pi)^{d}}c_{\mathbf{k}s}^{\dagger}\mathbf{v}{}_{\mathbf{k}ss'}c_{\mathbf{k}s'},\label{eq:J1}\\
\mathbf{J}_{A}(t) & =-e\sum_{ss'}\int\frac{d^{d}\mathbf{k}}{(2\pi)^{d}}c_{\mathbf{k}s}^{\dagger}\bigl[\mathbf{v}{}_{\mathbf{k}ss'}+\delta_{ss'}\frac{e}{m_{e}}\mathbf{A}(t)\bigr]c_{\mathbf{k}s'}.\label{eq:J2}
\end{align}
The expression of the current in the velocity gauge is a consequence
of Eqs.\,(\ref{eq:EQUIBANDS}) and (\ref{eq:BISCANON-1}). Any component
of velocity in operator form satisfies the general relation between
$\mathbf{k}$ diagonal observables in both gauges, 
\[
\mathcal{V}^{\alpha,A}\left(\mathbf{k}\right)=\mathcal{V}^{\alpha,E}\left(\mathbf{k}+\frac{e}{\hbar}\mathbf{A}\right).
\]
An expansion of the right hand side in powers of \textbf{A} produces
only two terms of orders $A^{(0)}$ and $A^{(1)}$. The remaining
terms involve two or more derivatives with respect to \textbf{k},
which, following Eq\@.\,(\ref{eq:COM3-1}), can be expressed in
terms of commutators of the covariant derivative and $\bigl[D_{\mathbf{k}}^{\beta},\mathcal{V}^{\alpha,E}(\mathbf{k})\bigr]$.
As the latter is proportional to the identity operator, Eq.\,(\ref{eq:BISCANON-1}),
the commutators are exactly zero and the higher order terms vanish. 

\begin{align}
\mathcal{V}^{\alpha,A}\left(\mathbf{k}\right)= & \mathcal{V}^{\alpha,E}\left(\mathbf{k}\right)+\frac{e}{\hbar}A^{\beta}\bigl[D_{\mathbf{k}}^{\beta},\mathcal{V}^{\alpha,E}(\mathbf{k})\bigr]\label{eq:v_operator_expansion}\\
= & \mathcal{V}^{\alpha,E}\left(\mathbf{k}\right)+\frac{e}{m}A^{\alpha}\hat{1}.
\end{align}
In matrix form this is 
\[
v_{\mathbf{k}ss'}^{\alpha,A}=v_{\mathbf{k}ss'}^{\alpha}+\delta_{ss'}\frac{e}{m_{e}}A^{\alpha}(t),
\]
where $\mathbf{v}{}_{\mathbf{k}ss'}$ is the velocity matrix in the
length gauge. 

\section{the reduced density matrix\label{sec:the-reduced-density}}

The reduced density matrix (RDM), is a matrix in band space, defined
by the average of momentum conserving inter band transitions ($s\ne s'$)
and the intra band ($s=s'$) transitions \footnote{Note the switch in the band indexes of the RDM.}.
\begin{equation}
\rho_{\mathbf{k}ss'}^{\alpha}(t)=\langle c_{\mathbf{k}s'}^{\dagger}c_{\mathbf{k}s}\rangle_{\alpha}.\label{eq:EQM1}
\end{equation}
The superscript $\alpha$ now denotes the gauge in which this object
is computed, $\alpha=A,E$. The average of any operator $\hat{A}$
is the trace 
\begin{equation}
\langle\hat{A}\rangle_{\alpha}=\textrm{tr}\bigl[\rho^{\alpha}\hat{A}\bigr],\label{eq:EQM2}
\end{equation}
where, $\rho^{\alpha}$, the full many-body density matrix is
\begin{equation}
\rho^{\alpha}=\sum_{n}p_{n}\left|\psi_{n}^{\alpha}\right\rangle \left\langle \psi_{n}^{\alpha}\right|,\label{eq:EQM3}
\end{equation}
and $\bigl\{\ket{\psi_{n}^{\alpha}}\bigr\}$ is a complete set of
state vectors. In the Schr\"{o}dinger picture, the time evolution
of $\rho^{\alpha}(t)$ is governed by the time-evolution of the state
vectors, $\left|\psi_{n}(t)\right\rangle $, 
\begin{equation}
i\hbar\frac{\partial\left|\psi_{n}^{\alpha}(t)\right\rangle }{\partial t}=H_{\alpha}(t)\left|\psi_{n}^{\alpha}(t)\right\rangle .\label{eq:EQM4}
\end{equation}
The equation of motion of the RDM takes the form 
\begin{eqnarray}
i\hbar\frac{\partial\rho_{\mathbf{k}ss'}^{\alpha}(t)}{\partial t} & = & \textrm{tr}\left[i\hbar\frac{\partial\rho^{\alpha}\left(t\right)}{\partial t}c_{\mathbf{k}s'}^{\dagger}c_{\mathbf{k}s}\right]\nonumber \\
 & = & \textrm{tr}\left[\bigl[H_{\alpha}(t),\rho^{\alpha}\left(t\right)\bigr]c_{\mathbf{k}s'}^{\dagger}c_{\mathbf{k}s}\right]\nonumber \\
 & = & \bigl\langle\bigl[c_{\mathbf{k}s'}^{\dagger}c_{\mathbf{k}s},H_{\alpha}(t)\bigr]\bigr\rangle_{\alpha}.\label{eq:EQM5}
\end{eqnarray}
The gauge freedom to express the uniform electric field is thereby
carried into the calculation of the system's dynamics, which can be
described either in terms of $\rho_{\mathbf{k}ss'}^{E}(t)$ or $\rho_{\mathbf{k}ss'}^{A}(t)$. 

Computing the commutators on the right hand side of Eq.\,(\ref{eq:EQM5}),
using the Hamiltonians, Eqs.\,(\ref{eq:HAMB1}) and (\ref{eq:HAMB2}),
we obtain closed equations of motion for the RDM \footnote{In order to simplify notation we will drop the crystal momentum label
from the objects inside the commutators and write it alongisde the
band indexes. },

\begin{align}
\bigl[i\hbar\frac{\partial}{\partial t}-\epsilon_{\mathbf{k}ss'}\bigr]\rho_{\mathbf{k}ss'}^{E}(t) & =ie\mathbf{E}(t)\cdot\left[\mathbf{D},\rho^{E}(t)\right]_{\mathbf{k}ss'},\label{eq:EQM8}\\
\bigl[i\hbar\frac{\partial}{\partial t}-\epsilon_{\mathbf{k}ss'}\bigr]\rho_{\mathbf{k}ss'}^{A}(t) & =e\mathbf{A}(t)\cdot\left[\mathbf{v},\rho^{A}(t)\right]_{\mathbf{k}ss'}.\label{eq:EQM9}
\end{align}
Here we have defined $\epsilon_{\mathbf{k}ss'}:=\epsilon_{\mathbf{k}s}-\epsilon_{\mathbf{k}s'}$.
The equation for the scalar potential RDM is found in references \cite{Aversa1995,Cheng2014,PhysRevB.91.235320},
but not, as here, cast in terms of the covariant derivative. The presence
of the derivative with respect to $\mathbf{k}$ in Eq.\,(\ref{eq:EQM8})
couples the response at different values of $\mathbf{k}$, whereas,
its counterpart for the vector potential gauge, Eq.\,(\ref{eq:EQM9}),
is completely decoupled in crystal momentum, \textbf{$\mathbf{k}$},
and can thus be solved independently for each point of the FBZ. Averages
of single particle observables, diagonal in momentum space, such as
the currents, (\ref{eq:J1}) and (\ref{eq:J2}), can be obtained from
the RDM's as traces over band space 
\begin{eqnarray}
\bigl\langle\mathbf{J}_{E}(t)\bigr\rangle & = & -e\int\frac{d^{d}\mathbf{k}}{(2\pi)^{d}}\text{Tr}\left[\boldsymbol{\mathcal{V}}^{E}(\mathbf{k})\rho^{E}(\mathbf{k},t)\right],\label{eq:CUR1}\\
\bigl\langle\mathbf{J}_{A}(t)\bigr\rangle & = & -e\int\frac{d^{d}\mathbf{k}}{(2\pi)^{d}}\text{Tr}\Bigl[\Bigl(\boldsymbol{\mathcal{V}}^{E}(\mathbf{k})+\frac{e}{m_{e}}\mathbf{A}(t)\hat{1}\Bigr)\,\rho^{A}(\mathbf{k},t)\Bigr],\nonumber \\
\label{eq:CUR2}
\end{eqnarray}
for $\hat{1}$ the identity in band space. Given that these two alternative
formulations are related by a unitary transformation, Eq.\,(\ref{eq:CUR1})
and Eq.\,(\ref{eq:CUR2}) have to yield the same results, although
this is far from obvious at this point. To relate these two objects,
we must first establish the relation between RDMs.

The full many-body density matrix $\rho^{A}$,
\begin{align}
\rho^{A} & =\sum_{n}p_{n}\ket{\psi_{n}^{A}(t)}\bra{\psi_{n}^{A}(t)},\label{eq:EQUI1}
\end{align}
can be expressed in the state vectors of its counterpart description
by means of a unitary transformation, Eq.\,(\ref{eq:UNITARY}), 
\begin{equation}
\rho^{A}=\mathcal{U}^{\dagger}(t)\,\rho^{E}\,\mathcal{U}(t).\label{eq:EQUI2}
\end{equation}
The vector potential RDM is therefore expressible as averages with
$\rho^{E}$, of suitably modified operators
\begin{eqnarray}
\rho_{\mathbf{k}ss'}^{A}(t) & = & \text{tr}\left[\rho^{A}c_{\mathbf{k}s'}^{\dagger}c_{\mathbf{k}s}\right]\nonumber \\
 & = & \text{tr}\left[\mathcal{U}^{\dagger}(t)\,\rho^{E}\,\mathcal{U}(t)c_{\mathbf{k}s'}^{\dagger}c_{\mathbf{k}s}\right]\nonumber \\
 & = & \text{tr}\left[\rho^{E}\,\mathcal{U}(t)c_{\mathbf{k}s'}^{\dagger}c_{\mathbf{k}s}\mathcal{U}^{\dagger}(t)\right]\nonumber \\
 & = & \text{tr}\left[\rho^{E}\,\tilde{c}_{\mathbf{k}s'}^{\dagger}\,\tilde{c}_{\mathbf{k}s}\right].\label{eq:EQUI3}
\end{eqnarray}
The new creation operator $\tilde{c}_{\mathbf{k}s'}^{\dagger}$ is
obtained from $c_{\mathbf{k}s'}^{\dagger}$ by the same unitary transformation
\begin{eqnarray}
\tilde{c}_{\mathbf{k}s'}^{\dagger} & = & \mathcal{U}(t)\,c_{\mathbf{k}s'}^{\dagger}\,\mathcal{U}^{\dagger}(t)\nonumber \\
 & = & \int d^{d}\mathbf{r}\,\Phi_{\mathbf{k}s'}(\mathbf{r})\,\Psi^{\dagger}(\mathbf{r})\label{eq:EQUI4}
\end{eqnarray}
 and creates an electron with wave function 
\begin{equation}
\Phi_{\mathbf{k}s'}(\mathbf{r}):=e^{ie\mathbf{r}\cdot\mathbf{A}(t)/\hbar}\psi_{\mathbf{k}s'}(\mathbf{r})=e^{i\mathbf{r}\cdot\left(\mathbf{k}+e\mathbf{A}(t)/\hbar\right)}u_{\mathbf{k}s'}(\mathbf{r}).\label{eq:EQUI5}
\end{equation}
While this is clearly a Bloch state with wave vector $\mathbf{q}=\mathbf{k}+e\mathbf{A}(t)/\hbar$,
\emph{it is not} $\psi_{\mathbf{k}+e\mathbf{A}(t)/\hbar\,s'}$, but
can be expanded as a linear combination of the shifted Bloch states,
\begin{eqnarray}
\left|\Phi_{\mathbf{k}s'}\right\rangle  & = & \sum_{r'}\int\frac{d^{d}\mathbf{q}}{(2\pi)^{d}}\bigl|\psi_{\mathbf{q}\,r'}\bigr\rangle\bigl\langle\psi_{\mathbf{q}r'}\bigl|\Phi_{\mathbf{k}s'}\bigr\rangle\nonumber \\
 & = & \sum_{r'}\,\bigl|\psi_{\mathbf{k}+e\mathbf{A}(t)/\hbar\,r'}\bigr\rangle\bigl\langle u_{\mathbf{k}+e\mathbf{A}(t)/\hbar\,r'}\bigl|u_{\mathbf{k}s'}\bigr\rangle.\label{eq:EQUI6}
\end{eqnarray}
This allows us to relate $\tilde{c}_{\mathbf{k}s'}^{\dagger}$ and
$\tilde{c}_{\mathbf{k}s'}$ to the original creation and destruction
operators
\begin{eqnarray}
\tilde{c}_{\mathbf{k}s'}^{\dagger} & = & \sum_{r'}\,\bigl\langle u_{\mathbf{k}+e\mathbf{A}(t)/\hbar r'}\bigl|u_{\mathbf{k}s'}\bigr\rangle\,c_{\mathbf{k}+e\mathbf{A}(t)/\hbar\,r'}^{\dagger},\label{eq:EQUI7}
\end{eqnarray}
and express the RHS of Eq.\,(\ref{eq:EQUI3}) in terms of the shifted
Bloch operators. Since 
\begin{eqnarray}
\text{tr}\bigl[\rho^{E}c_{\mathbf{k}+e\mathbf{A}(t)/\hbar\,r'}^{\dagger}c_{\mathbf{k}+e\mathbf{A}(t)/\hbar\,r}\bigr] & = & \langle c_{\mathbf{k}+e\mathbf{A}(t)/\hbar\,r'}^{\dagger}c_{\mathbf{k}+e\mathbf{A}(t)/\hbar\,r}\rangle_{E}\nonumber \\
 & = & \rho_{\mathbf{k}+e\mathbf{A}(t)/\hbar\,rr'}^{E}(t),\label{eq:EQUI8}
\end{eqnarray}
using this and Eq.\,(\ref{eq:EQUI7}) in Eq.\,(\ref{eq:EQUI3})
we obtain the relation between $\rho_{ss'}^{A}(\mathbf{k},t)$ and
$\rho_{ss'}^{E}(\mathbf{k},t)$ as\begin{widetext}
\begin{eqnarray}
\rho_{\mathbf{k}ss'}^{A}(t) & = & \sum_{rr'}\braket{u_{\mathbf{k}s}}{u_{\mathbf{k}+e\mathbf{A}(t)/\hbar\,r'}}\rho_{\mathbf{k}+e\mathbf{A}(t)/\hbar\,rr'}^{E}(t)\braket{u_{\mathbf{k}+e\mathbf{A}(t)/\hbar\,r}}{u_{\mathbf{k}s'}}.\label{eq:EQUI9}
\end{eqnarray}
\end{widetext}Following our definition of operators in band space,
Eq.\,(\ref{eq:BISMAT-1}), this equality can be cast in operator
form\footnote{The hats will be henceforth dropped.}, 
\begin{eqnarray}
\hat{\rho}^{A}(\mathbf{k},t) & = & \hat{\rho}^{E}\left(\mathbf{k}+\frac{e}{\hbar}\mathbf{A}(t),t\right).\label{eq:EQUI10}
\end{eqnarray}
The simplicity of this relation between the density matrices in the
velocity and length gauges, which amounts to a simple shift in the
value of $\mathbf{k}$, is only true in the operator representation;
as can be seen in Eq.\,(\ref{eq:EQUI9}), the matrix elements of
these two operators between states of a single basis $\left\{ \ket{u_{\mathbf{k}s}},\,s=0,1,\dots\right\} $
\emph{with the same}\textbf{\emph{ }}\textbf{$\mathbf{k}$}, do not
satisfy this simple relation. 

We can now show that the expectation values of observables in the
two descriptions are exactly the same. Consider,
\begin{align}
\langle\mathcal{O}^{A}(t)\rangle & =\int\frac{d^{d}\mathbf{k}}{\left(2\pi\right)^{d}}\text{Tr}\left[\mathcal{O}^{A}(\mathbf{k})\rho^{A}(\mathbf{k},t)\right].\label{eq:EQUI11}
\end{align}
Both the RDM and the observable operators are equal to their scalar
potential counterparts when their argument is adequately translated,
Eqs.\,(\ref{eq:EQUIBANDS}) and (\ref{eq:EQUI10}), 
\begin{equation}
\langle\mathcal{O}^{A}(t)\rangle=\int\frac{d^{d}\mathbf{k}}{\left(2\pi\right)^{d}}\text{Tr}\left[\mathcal{O}^{E}\left(\mathbf{k}+\frac{e}{\hbar}\mathbf{A}\right)\rho^{E}\left(\mathbf{k}+\frac{e}{\hbar}\mathbf{A}\right)\right].\label{eq:SHIFT_between_Gauges}
\end{equation}
This shifts the FBZ by a constant, which is irrelevant as the integrand
is periodic. This means that, 
\begin{align}
\langle\mathcal{O}^{A}(t)\rangle & =\int\frac{d^{d}\mathbf{k}}{\left(2\pi\right)^{d}}\text{Tr}\left[\mathcal{O}^{E}(\mathbf{k})\rho^{E}(\mathbf{k},t)\right]\nonumber \\
 & =\langle\mathcal{O}^{E}(t)\rangle.
\end{align}
The sum rules in reference \cite{Aversa1995} can be traced back to
this equivalence (see Appendix A). 

\section{Solutions to the rdm equations of motion}

A system's current response (in either scalar or vector potential
formalism) is obtained from the solution to its respective RDM equation
of motion. To write these solutions explicitly, we must employ the
general procedure from nonlinear physics: break the RDM into contributions
of different powers on the external field, $\rho=\rho^{(0)}+\rho^{(1)}+(...)$
and then proceed to iteratively solve the equations of motion for
each order. For $\rho^{(n)}$, these read as

\begin{eqnarray}
\left[i\hbar\frac{\partial}{\partial t}-\epsilon_{\mathbf{k}ss'}\right]\rho_{\mathbf{k}ss'}^{(n),E}(t) & = & ie\mathbf{E}(t)\cdot\bigl[\mathbf{D},\rho^{(n-1),E}(t)\bigr]_{\mathbf{k}ss'},\nonumber \\
\label{eq:EQMOTN2}
\end{eqnarray}
\begin{eqnarray}
\left[i\hbar\frac{\partial}{\partial t}-\epsilon_{\mathbf{k}ss'}\right]\rho_{\mathbf{k}ss'}^{(n),A}(t) & = & e\mathbf{A}(t)\cdot\bigl[\mathbf{v},\rho^{(n-1),A}(t)\bigr]_{\mathbf{k}ss'}.\nonumber \\
\label{eq:EQMOTN1}
\end{eqnarray}
In expressing the time-dependent objects in terms of their Fourier
decomposition, we assume adiabatic switching of the perturbation ($\eta\rightarrow0^{+}$),
\begin{eqnarray}
\rho_{\mathbf{k}ss'}^{(n),\alpha}(t) & = & \int\frac{d\omega}{2\pi}\,e^{-i\left(\omega+i\eta\right)t}\,\rho_{\mathbf{k}ss'}^{(n),\alpha}(\omega),\\
\mathbf{E}(t) & = & \int\frac{d\omega}{2\pi}\,e^{-i\left(\omega+i\eta\right)t}\,\mathbf{E}(\omega),\\
\mathbf{A}(t) & = & \int\frac{d\omega}{2\pi}\,e^{-i\left(\omega+i\eta\right)t}\,\mathbf{A}(\omega).
\end{eqnarray}
The time derivative in the equations of motion is replaced by a frequency
factor that is collected into an energy denominator,
\begin{eqnarray}
d_{\mathbf{k}ss'}(\omega) & := & \frac{1}{\hbar\omega-\epsilon_{\mathbf{k}ss'}}.
\end{eqnarray}
From this point on, the frequency argument of these energy denominators
is understood to have an infinitesimal imaginary part. Using the Hadamard
product of two matrices, 
\begin{eqnarray}
\left(A\circ B\right)_{ss'} & := & A_{ss'}\,B_{ss'},
\end{eqnarray}
we write recursion relations for the RDM solutions, 
\begin{eqnarray}
\rho_{\mathbf{k}ss'}^{(n),E}(\omega) & = & ie\int\frac{d\omega_{1}}{2\pi}\,E^{\alpha_{1}}(\omega_{1})\,(d(\omega)\nonumber \\
 &  & \circ\bigl[D^{\alpha_{1}},\rho^{(n-1),E}(\omega-\omega_{1})\bigr])_{\mathbf{k}ss'},
\end{eqnarray}
\begin{eqnarray}
\rho_{\mathbf{k}ss'}^{(n),A}(\omega) & = & e\int\frac{d\omega_{1}}{2\pi}\,A^{\alpha_{1}}(\omega_{1})\,(d(\omega)\nonumber \\
 &  & \circ\bigl[v^{\alpha_{1}},\rho^{(n-1),A}(\omega-\omega_{1})\bigr])_{\mathbf{k}ss'}.
\end{eqnarray}
The successive application of these expressions, brings the $n$-th
order solution, $\rho_{\mathbf{k}ss'}^{(n)}(\omega)$, to the form
of nested commutators of the zeroth order one, $\rho_{\mathbf{k}ss'}^{(0)}$,
which is the Fermi-Dirac distribution function times the unit matrix
in band space, $\rho_{\mathbf{k}ss'}^{(0)}=f_{\mathbf{k}s}\delta_{ss'}$.
With one last bit of notation, 
\begin{align}
\omega_{\left[m\right]} & :=\sum_{i=1}^{m}\omega_{i},
\end{align}
we write $\rho_{\mathbf{k}ss'}^{(n)}$ in each formalism as\begin{widetext}
\begin{eqnarray}
\rho_{\mathbf{k}ss'}^{(n),E}(\omega) & = & \left(ie\right)^{n}\left[\prod_{i=1}^{n-1}\int\frac{d\omega_{i}}{2\pi}\,E^{\alpha_{i}}(\omega_{i})\right]E^{\alpha_{n}}(\omega-\omega_{\left[n-1\right]})\,\left(d(\omega)\circ\bigl[D^{\alpha_{1}},\,d(\omega-\omega_{1})\circ\bigl[...,d(\omega-\omega_{\left[n-1\right]})\circ\bigl[D^{\alpha_{n}},\rho^{(0)}\bigr]...\bigr]\bigr]\right)_{\mathbf{k}ss'},\nonumber \\
\label{eq:RECURSIVE1}
\end{eqnarray}
\begin{eqnarray}
\rho_{\mathbf{k}ss'}^{(n),A}(\omega) & = & e^{n}\left[\prod_{i=1}^{n-1}\int\frac{d\omega_{i}}{2\pi}\,A^{\alpha_{i}}(\omega_{i})\right]A^{\alpha_{n}}(\omega-\omega_{\left[n-1\right]})\,\left(d(\omega)\circ\bigl[v^{\alpha_{1}},\,d(\omega-\omega_{1})\circ\bigl[...,d(\omega-\omega_{\left[n-1\right]})\circ\bigl[v^{\alpha_{n}},\rho^{(0)}\bigr]...\bigr]\bigr]\right)_{\mathbf{k}ss'}.\nonumber \\
\label{eq:RECURSIVE2}
\end{eqnarray}
\end{widetext}The use of the nested commutator and the factor $d(\omega)$
allow for a compact form of these solutions (albeit hiding their considerable
complexity). These solutions are entirely written in terms of three
objects (and their derivatives, in the case of $\rho^{E}$): the band
energies, $\epsilon_{\mathbf{k}s}$; the Berry connection, $\xi_{\mathbf{k}ss'}$;
and the structure of filled/empty bands, $f_{\mathbf{k}s}$. In principle,
by determining these, one determines a system's nonlinear response,
provided that one can compute the FBZ integrals in Eqs.(\ref{eq:CUR1})
and (\ref{eq:CUR2}). In section \ref{sec:Effective-hamiltonians},
we use these expressions to discuss whether it is possible to truncate
the sums over bands, implicit in these matrix products, to a reduced
set. But before, we study the first two non-trivial orders of the
monolayer graphene response to a uniform electric field, to illustrate
that these reproduce the results of references \cite{Cheng2014,PhysRevB.91.235320,Mikhailov2016}.

\section{Linear and third order response in the monolayer graphene \label{sec:Linear-and-third}}

Monolayer graphene is (usually) described by a tight-binding model
that considers only nearest neighbour hopping\cite{Peres2009} and
for which the expressions for the two bands, $\epsilon_{\mathbf{k}s}$,
and the periodic functions, $u_{\mathbf{k}s}$, can be explicitly
computed. This allows one to derive some useful properties.

First, a double band-index sum, such as the ones in $\langle\mathbf{J}_{E}\rangle$
and $\langle\mathbf{J}_{A}\rangle$, over an antisymmetric object
$\theta_{ss'}=-\theta_{s's}$, can be reduced to a single sum, where
$\bar{s}$ reads as the band opposite to $s$,
\begin{align}
\sum_{s's}\theta_{ss'} & \rightarrow\sum_{s}\theta_{s\bar{s}},\label{eq:TRACE}
\end{align}
Second, the Berry connection for the monolayer is described by an
intra band and an inter band term. With adequate choice of gauge (Eq.~\ref{eq:gauge_k_space})
the following properties can be obtained \cite{Cheng2014,PhysRevB.91.235320,Mikhailov2016},
\begin{align}
\xi_{\mathbf{k}ss}^{\alpha} & =\xi_{\mathbf{k}\bar{s}\bar{s}}^{\alpha},\label{eq:BC1}\\
\xi_{\mathbf{k}s\bar{s}}^{\alpha} & =\xi_{\mathbf{k}\bar{s}s}^{\alpha}.\label{eq:BC2}
\end{align}
In addition, we can choose these to be even under $\mathbf{k}\rightarrow-\mathbf{k}$.

\subsection*{Linear order response}

The first order term of the RDM is 
\begin{eqnarray}
\rho_{\mathbf{k}ss'}^{(1)}(\omega) & = & ie\,E^{\alpha}(\omega)\,\left(d(\omega)\circ\bigl[D^{\alpha},\rho^{(0)}\bigr]\right)_{\mathbf{k}ss'},\\
 & = & ie\,E^{\alpha}(\omega)\,\Bigl\{\delta_{ss'}\frac{1}{\hbar\omega}\nabla_{\mathbf{k}}^{\alpha}f_{\mathbf{k}s}-i\frac{\xi_{\mathbf{k}ss'}^{\alpha}\left(f_{\mathbf{k}s'}-f_{\mathbf{k}s}\right)}{\hbar\omega-\epsilon_{\mathbf{k}ss'}}\Bigr\}.\nonumber \\
\end{eqnarray}
Writing the current, Eq.\,(\ref{eq:CUR1}), in terms of its Fourier
components, we obtain two contributions \cite{Cheng2014,PhysRevB.91.235320,Mikhailov2016},
\begin{eqnarray}
\langle J^{(1),\beta}(\omega)\rangle & = & E^{\alpha}(\omega)\int\frac{d^{d}\mathbf{k}}{(2\pi)^{d}}\bigl[\Pi_{\mathbf{k},\textrm{intra}}^{(1),\beta\alpha}(\omega)+\Pi_{\mathbf{k},\textrm{inter}}^{(1),\beta\alpha}(\omega)\bigr].\nonumber \\
\end{eqnarray}
The intra band contribution is a generalized Drude term,
\begin{eqnarray}
\Pi_{\mathbf{k},\textrm{intra}}^{(1),\beta\alpha}(\omega) & = & i\frac{e^{2}}{\hbar^{2}}\frac{1}{\omega}\sum_{s}\bigl(\nabla_{\mathbf{k}}^{\beta}\epsilon_{\mathbf{k}s}\bigr)\bigl(\nabla_{\mathbf{k}}^{\alpha}\epsilon_{\mathbf{k}s}\bigr)\left(-\frac{\partial f_{\mathbf{k}s}}{\partial\epsilon_{\mathbf{k}s}}\right).\nonumber \\
\end{eqnarray}
 and the inter band contribution, $\Pi_{\textrm{inter}}^{\alpha\beta,(1)}(\mathbf{k},\omega)$,
which involves the Berry connection,
\begin{eqnarray}
\Pi_{\mathbf{k},\textrm{inter}}^{(1),\beta\alpha}(\omega) & = & -e^{2}\sum_{s}v_{\mathbf{k},s\bar{s}}^{\beta}\,\xi_{\mathbf{k},\bar{s}s}^{\alpha}\,\frac{f_{\mathbf{k}s}-f_{\mathbf{k}\bar{s}}}{\hbar\omega-\epsilon_{\mathbf{k}\bar{s}s}}.\nonumber \\
\end{eqnarray}

\subsection*{Third order response}

Because graphene has inversion symmetry, its second order response
is zero (see Appendix\,\ref{sec:Current-response-in}). The first
nonlinear contribution to the current is thus the third order one,
which is to be computed here. From the general RDM solution, Eq.\,(\ref{eq:RECURSIVE1}),
we obtain for $n=3$,\begin{widetext}
\begin{eqnarray}
\rho_{\mathbf{k}ss'}^{(3)}(\omega) & = & \left(ie\right)^{3}\int\frac{d\omega_{1}}{2\pi}\int\frac{d\omega_{2}}{2\pi}\,E^{\alpha_{1}}(\omega_{1})\,E^{\alpha_{2}}(\omega_{2})\,E^{\alpha_{3}}(\omega-\omega_{[2]})\,\Bigl(d(\omega)\circ\bigl[D^{\alpha_{1}},\,d(\omega-\omega_{1})\circ\bigl[D^{\alpha_{2}},d(\omega-\omega_{[2]})\circ\bigl[D^{\alpha_{3}},\rho^{(0)}\bigr]\bigr]\bigr]\Bigr)_{\mathbf{k}ss'}.\nonumber \\
\label{eq:RHO3}
\end{eqnarray}
\end{widetext} Expanding this by means of Eq.\,(\ref{eq:COM1-1}),
produces a plethora of terms which we organize, following Mikhailov,
by the number of intra band (derivatives) and inter band (Berry connections)
factors. These can be manipulated independently and give contributions
to the current, that can be labeled by $i=1,2,3$, the number of intra
band factors.

The field factors and the FBZ integrations are the same in all these
contributions, and so \begin{widetext}
\begin{eqnarray}
\langle J^{(3),\beta}(\omega)\rangle & = & e^{4}\int\frac{d\omega_{1}}{2\pi}\int\frac{d\omega_{2}}{2\pi}\,E^{\alpha_{1}}(\omega_{1})\,E^{\alpha_{2}}(\omega_{2})\,E^{\alpha_{3}}(\omega-\omega_{[2]})\int\frac{d^{d}\mathbf{k}}{(2\pi)^{d}}\sum_{i=0}^{3}\Pi_{\mathbf{k},i}^{(3),\beta\alpha_{1}\alpha_{2}\alpha_{3}}(\omega,\omega_{1},\omega_{2}).
\end{eqnarray}
 \end{widetext}The $\Pi_{3}$ contribution is the single term in
Eq.\,(\ref{eq:RHO3}) with three \textbf{$\mathbf{k}$}-derivatives,
while $\Pi_{2}$ gathers three terms with two $\mathbf{k}$-derivatives,
to which we can apply the property Eq.\,(\ref{eq:TRACE}). Contributions
$\Pi_{1}$ and $\Pi_{0}$ require additional manipulations, presented
in Appendix C. Consistently with our notation $\epsilon_{\mathbf{k}s\bar{s}}:=\epsilon_{\mathbf{k}s}-\epsilon_{\mathbf{k}\bar{s}},$
we abbreviate the factor $f_{\mathbf{k}s}-f_{\mathbf{k}\bar{s}}$
as $f_{\mathbf{k}s\bar{s}}$. The expressions for the $\Pi_{i}$ are:
\begin{widetext}
\begin{eqnarray}
\Pi_{3}^{(3),\beta\alpha_{1}\alpha_{2}\alpha_{3}} & = & \frac{i}{\hbar\omega}\frac{1}{\hbar(\omega-\omega_{1})}\frac{1}{\hbar(\omega-\omega_{[2]})}\sum_{s}v_{\mathbf{k}ss}^{\beta}\nabla_{\mathbf{k}}^{\alpha_{1}}\nabla_{\mathbf{k}}^{\alpha_{2}}\nabla_{\mathbf{k}}^{\alpha_{3}}f_{\mathbf{k}s},\label{eq:3}
\end{eqnarray}
\begin{eqnarray}
\Pi_{2}^{(3),\beta\alpha_{1}\alpha_{2}\alpha_{3}} & = & \sum_{s}v_{\mathbf{k}\bar{s}s}^{\beta}\frac{1}{\hbar\omega-\epsilon_{\mathbf{k}s\bar{s}}}\biggl\{\frac{1}{\hbar(\omega-\omega_{1})}\frac{1}{\hbar(\omega-\omega_{[2]})}\,\xi_{\mathbf{k}s\bar{s}}^{\alpha_{1}}\,\nabla_{\mathbf{k}}^{\alpha_{2}}\nabla_{\mathbf{k}}^{\alpha_{3}}f_{\mathbf{k}\bar{ss}}\nonumber \\
 &  & +\frac{1}{\hbar(\omega-\omega_{[2]})}\nabla_{\mathbf{k}}^{\alpha_{1}}\left(\frac{\xi_{\mathbf{k}s\bar{s}}^{\alpha_{2}}\nabla_{\mathbf{k}}^{\alpha_{3}}f_{\mathbf{k}\bar{ss}}}{\hbar(\omega-\omega_{1})-\epsilon_{\mathbf{k}s\bar{s}}}\right)\\
 &  & +\nabla_{\mathbf{k}}^{\alpha_{1}}\left(\frac{1}{\hbar(\omega-\omega_{1})-\epsilon_{\mathbf{k}s\bar{s}}}\nabla_{\mathbf{k}}^{\alpha_{2}}\left(\frac{\xi_{\mathbf{k}s\bar{s}}^{\alpha_{3}}\,f_{\mathbf{k}\bar{ss}}}{\hbar(\omega-\omega_{[2]})-\epsilon_{\mathbf{k}s\bar{s}}}\right)\right)\biggr\},\label{eq:2}
\end{eqnarray}
\begin{eqnarray}
\Pi_{1}^{(3),\beta\alpha_{1}\alpha_{2}\alpha_{3}} & = & \frac{i}{\hbar\omega}\sum_{s}v_{\mathbf{k}ss}^{\beta}\biggl\{\frac{1}{\hbar(\omega-\omega_{1})}\nabla_{\mathbf{k}}^{\alpha_{1}}\left(\xi_{\mathbf{k}s\bar{s}}^{\alpha_{2}}\xi_{\mathbf{k}\bar{s}s}^{\alpha_{3}}\,f_{\mathbf{k}s\bar{s}}\left(\frac{1}{\hbar(\omega-\omega_{[2]})-\epsilon_{\mathbf{k}\bar{s}s}}+\frac{1}{\hbar(\omega-\omega_{[2]})-\epsilon_{\mathbf{k}\bar{s}s}}\right)\right)\nonumber \\
 &  & +\frac{\xi_{\mathbf{k}s\bar{s}}^{\alpha_{1}}}{\hbar(\omega-\omega_{1})-\epsilon_{\mathbf{k}\bar{s}s}}\nabla_{\mathbf{k}}^{\alpha_{2}}\left(\xi_{\mathbf{k}\bar{s}s}^{\alpha_{3}}f_{\mathbf{k}s\bar{s}}\left(\frac{1}{\hbar(\omega-\omega_{[2]})-\epsilon_{\mathbf{k}\bar{s}s}}+\frac{1}{\hbar(\omega-\omega_{[2]})-\epsilon_{\mathbf{k}s\bar{s}}}\right)\right)\nonumber \\
 &  & +\frac{1}{\hbar(\omega-\omega_{[2]})}\xi_{\mathbf{k}s\bar{s}}^{\alpha_{1}}\xi_{\mathbf{k}\bar{s}s}^{\alpha_{2}}\left(\frac{1}{\hbar(\omega-\omega_{1})-\epsilon_{\mathbf{k}\bar{s}s}}+\frac{1}{\hbar(\omega-\omega_{1})-\epsilon_{\mathbf{k}s\bar{s}}}\right)\nabla_{\mathbf{k}}^{\alpha_{3}}f_{\mathbf{k}s\bar{s}}\biggr\},\label{eq:1}
\end{eqnarray}
\begin{align}
\Pi_{0}^{(3),\beta\alpha_{1}\alpha_{2}\alpha_{3}} & =\frac{2}{\hbar(\omega-\omega_{1})}\sum_{s}\frac{1}{\hbar\omega-\epsilon_{\mathbf{k}s\bar{s}}}v_{\mathbf{k}\bar{s}s}^{\beta}\,\xi_{\mathbf{k}s\bar{s}}^{\alpha_{1}}\,\xi_{\mathbf{k}\bar{s}s}^{\alpha_{2}}\,\xi_{\mathbf{k}s\bar{s}}^{\alpha_{3}}\,f_{\mathbf{k}\bar{s}s}\left(\frac{1}{\hbar(\omega-\omega_{[2]})-\epsilon_{\mathbf{k}s\bar{s}}}+\frac{1}{\hbar(\omega-\omega_{[2]})-\epsilon_{\mathbf{k}\bar{s}s}}\right).\label{eq:0}
\end{align}
\end{widetext} Apart from differences in the Cartesian and frequency
indexes, which can always be relabeled, these are the expressions
found in \cite{Cheng2014,PhysRevB.91.235320,Mikhailov2016}, in the
limit where $\Gamma_{(i)}$, $\Gamma_{(e)}$, the phenomenological
scattering rates, are set to zero.

\section{Effective hamiltonians\label{sec:Effective-hamiltonians}}

As we have seen in the previous sections, the current response in
the Schr\"{o}dinger problem can be computed by two different but
equivalent procedures. Although this is conceptually important, one
is ultimately interested in computing the current in materials described
by effective Hamiltonians, that account only for a finite number of
bands. 

We will now show that, by truncating the band space, the scalar and
vector potential currents are no longer the same; the latter contains
relevant contributions from bands that are left out of the effective
Hamiltonian.

\begin{figure}
\includegraphics[scale=0.6]{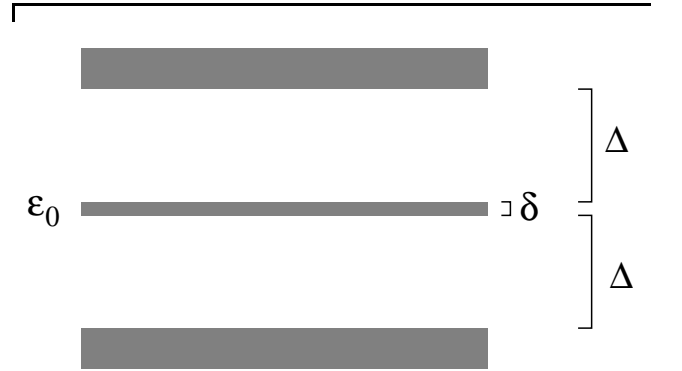}

\caption{A conceptual picture of an effective Hamiltonian that describes the
bands in subspace $\mathcal{E}_{0}$. The Fermi level lies somewhere
in that subspace. For bands inside $\mathcal{E}_{0}$ , the energy
difference is of order $\delta$, while the energy difference between
bands in different subspaces is of order $\Delta\gg\delta$. \label{fig:A-conceptual-image}}
\end{figure}

In Fig.\,(\ref{fig:A-conceptual-image}), we present a conceptual
picture of a spectrum which has a cluster of bands close to the Fermi
level, which we deem relevant, well separated in energy by the bands
below (filled) and above (empty). We denote the energy scale in the
subspace $\mathcal{E}_{0}$ of relevant bands by $\delta$ and the
energy separation to other bands by $\Delta\gg\delta$; we assume
the frequency of the external field, $\omega$, to be of the order,
$\omega\sim\delta/\hbar$. The question we wish to answer is whether
the bands outside $\mathcal{E}_{0}$ can be ignored in the calculation
of the current. 

Energy denominators, $d(\omega)$, involving bands inside $\mathcal{E}_{0}$,
\[
d_{\mathbf{k}ss'}(\omega)=\frac{1}{\hbar\omega-\epsilon_{\mathbf{k}ss'}},\qquad\epsilon_{\mathbf{k}ss'}\sim\delta,
\]
will be larger then those involving transition to and from bands in
$\mathcal{E}_{0}$ and those outside 
\[
d_{\mathbf{k}ss'}(\omega)=\frac{1}{\hbar\omega-\epsilon_{\mathbf{k}ss'}},\qquad\epsilon_{\mathbf{k}ss'}\sim\Delta.
\]
In the $n$-th order contribution to the current in the scalar potential
gauge, each term is a trace of a product $n+1$ matrices in band space.
\begin{equation}
\sum_{s's}v_{\mathbf{k}s's}^{\beta}d_{\mathbf{k}ss'}(\omega)\bigl[D^{\alpha_{1}},\,d(\omega-\omega_{1})\circ\bigl[...\bigl[D^{\alpha_{n}},\rho^{(0)}\bigr]...\bigr]\bigr]_{\mathbf{k}ss'}.\label{eq:EFFE3}
\end{equation}
One such term is the fully intra band one, in which we pick only the
diagonal part of each covariant derivative operator (see an example
in Eq.\,(\ref{eq:3})). This term is non zero only for the band that
contains the Fermi level; it has no contributions from bands outside
$\mathcal{E}_{0}.$ 

All inter band contributions contain at least one difference of occupation
factors, thus allowing us to discard any terms that involve only filled,
or only empty bands. What remains are contributions of three types:
(a) terms that involve transition between filled and empty bands,
both outside $\mathcal{E}_{0}$; (b) terms that involve transition
between bands in $\mathcal{E}_{0}$ and bands outside $\mathcal{E}_{0}$;
(c) inter band terms among the bands in $\mathcal{E}_{0}$. 

In Eq.\,(\ref{eq:EFFE3}), if $s'$ belongs to a filled band outside
$\mathcal{E}_{0}$, and $s$ to a empty band, also outside $\mathcal{E}_{0}$
(or vice-versa), $\epsilon_{\mathbf{k}ss'}\sim\Delta\gg\hbar\omega$
and
\begin{align*}
v_{\mathbf{k}s's}^{\beta}d_{\mathbf{k}ss'}(\omega)\sim\frac{i}{\hbar}\frac{\epsilon_{\mathbf{k}ss'}\xi_{\mathbf{k}s's}^{\beta}}{\hbar\omega-\epsilon_{\mathbf{k}ss'}} & =-\frac{i}{\hbar}\xi_{\mathbf{k}s's}^{\beta};
\end{align*}
the matrix that follows this term in Eq.\,(\ref{eq:EFFE3}), must
include at least an energy denominator $\sim\mathcal{O}(1/\Delta)$
because the last band index is the same as the first, $s'$. If $s'$
and $s$ both refer to a filled band (or an empty one) outside $\mathcal{E}_{0}$
the inter band terms between filled and empty bands must have at least
two such energy denominators. In other words, terms of type (a) have
a least and energy denominator of order $\sim\mathcal{O}(1/\Delta).$
An identical argument can be made for transitions between bands inside
and and outside $\mathcal{E}_{0}$, i.e. for transitions of type (b).
We conclude that, in the limit of $\Delta\gg\delta$, the dominant
terms come from bands in $\mathcal{E}_{0}$ (terms of type (c)) and
truncation to this subspace is a valid approximation. 

The same argument does not carry to the corresponding contribution
to the current in the velocity gauge:
\begin{equation}
\sum_{s's}v_{\mathbf{k}s's}^{\beta}d_{\mathbf{k}ss'}(\omega)\bigl[v^{\alpha_{1}},\,d(\omega-\omega_{1})\circ\bigl[...\bigl[v^{\alpha_{n}},\rho^{(0)}\bigr]...\bigr]\bigr]_{\mathbf{k}ss'}.\label{eq:EFFE4}
\end{equation}
In this case, every energy denominator is associated with a velocity
matrix element. In any transition involving energies $\epsilon_{rr'}\sim\mathcal{O}(\Delta)$,
the $d_{rr'}(\omega)\sim\mathcal{O}(1/\Delta)$, as before, but the
corresponding velocity matrix element has an off-diagonal contribution
$v_{\mathbf{k}r'r}^{\beta}\sim\xi_{\mathbf{k}rr'}^{\beta}\mathcal{O}(\Delta)$,
and such terms give relevant contributions no matter how large $\Delta$
is. In other words, bands away from the Fermi surface contribute just
as much as those in $\mathcal{E}_{0}$ for the expression of the current
in the velocity gauge.

\section{Summary and conclusions}

The concept of covariant derivative in $\mathbf{k}$-space has been
shown here to be of considerable value in the calculation of the nonlinear
current response. It is a very convenient representation of the position
operator {[}Eq.\,(\ref{eq:postion_cov_derivatiove}){]}; it clarifies
the structure of the velocity matrix {[}Eq.\,(\ref{eq:VELCOM}){]};
it allows a complete parallel development of the structure of the
reduced density matrices (RDM) in the length (scalar potential) and
velocity (vector potential) gauges {[}Eqs.\,(\ref{eq:EQM8}) and\,(\ref{eq:EQM9}){]}
and it provides compact expressions for the perturbative solutions
of the equations of motion of the RDM {[}Eqs.\,(\ref{eq:RECURSIVE1})
and\,(\ref{eq:RECURSIVE2}){]}. It also allowed us to see how the
equivalence between objects in the two gauges breaks down when the
band space is truncated.

Furthermore that the approximation is legitimate in the length gauge,
but fails in the velocity one is made clear by considering the truncation
of the covariant derivative and the velocity operators to a restricted
set of bands. The commutator of these two quantities, Eq.\,(\ref{eq:BISCANON-1}),
which is constant for the case of the infinite bands of the Schr\"{o}dinger
Hamiltonian, no longer holds for a truncated subset of these. In fact,
the statement that the commutator is constant is equivalent to the
Hamiltonian being linear or quadratic in \textbf{k} (the former is
the case of the Dirac Hamiltonian, where the commutator gives zero).
As a result, Eq.\,(\ref{eq:v_operator_expansion}) is no longer valid
for a general effective Hamiltonian and there is no equality between
currents in the two gauges, at least as they are written in Eqs.\,(\ref{eq:J1})
and (\ref{eq:J2}). In order to use the velocity gauge in actual calculations,
one must start from the beginning with the effective Hamiltonian and
perform the minimum coupling then. Naturally, this means modifying
both the equation of motion of the RDM, Eq.\,(\ref{eq:EQM9}), and
the current operator in the velocity gauge. This will be the subject
of a future paper.

\section{Acknowledgments}

The work of G.B.V and D.J.P is supported by Funda\c{c}\~{a}o para
a Ci\^{e}ncia e Tecnologia (FCT) under the grants PD/BI/129220/2017
and PD/BD/135019/2017 respectively. N.M.R.P. acknowledges funding
from the European Commission within the project \textquotedblleft Graphene-Driven
Revolutions in ICT and Beyond\textquotedblright{} (ref. no. 696656)
and the Portuguese Foundation for Science and Technology (FCT) in
the framework of the Strategic Financing UID/FIS/04650/2013.

\appendix

\section{The Aversa and Sipe Sum rules}

In reference \cite{Aversa1995}, Aversa and Sipe sketch how problems
might arise in the velocity gauge as it involves additional terms
(with respect to the length gauge) that are zero only when they are
treated exactly. They show this explicitly in the linear response. 

In our formalism this can be seen in full generality, starting from
Eqs.\,(\ref{eq:EQUI13}) and (\ref{eq:SHIFT_between_Gauges}). Expanding
in powers of \textbf{$\mathbf{A}(t)$}, we can write $\mathcal{\langle O}^{A}\left(\mathbf{k},t\right)\rangle$
in terms of $\langle\mathcal{O}^{E}\left(\mathbf{k},t\right)\rangle$
and contributions that depend explicitly on $\mathbf{A}(t)$, that
is, the $n\geq1$ terms of the sum,
\begin{eqnarray}
\mathcal{\langle O}^{A}\left(\mathbf{k},t\right)\rangle & = & \langle\mathcal{O}^{E}\left(\mathbf{k},t\right)\rangle\nonumber \\
 &  & +\frac{e}{\hbar}A^{\alpha_{1}}(t)\int\frac{d^{d}\mathbf{k}}{\left(2\pi\right)^{d}}\mathrm{Tr}\bigl[D^{\alpha_{1}},\mathcal{O}^{E}\left(\mathbf{k},t\right)\rho^{E}(\mathbf{k},t)\bigr]\nonumber \\
 &  & +(...),\label{eq:AVERSA_SUM}
\end{eqnarray}
The equivalence of the two gauges requires that everything other than
$\langle\mathcal{O}^{E}\left(\mathbf{k},t\right)\rangle$ in the right
hand side to be zero. Each contribution contains a factor which, for
arbitrary order $n$, reads as
\begin{equation}
\int\frac{d^{d}\mathbf{k}}{\left(2\pi\right)^{d}}\mathrm{Tr}\bigl[D^{\alpha_{1}},\mathrm{G}(\mathbf{k},t)\bigr],\label{eq:SUM_RULE_TERM}
\end{equation}
for $\mathrm{G}(\mathbf{k},t)$ is some matrix in band space, 
\[
\mathrm{G}(\mathbf{k},t):=\left[D^{\alpha_{2}},...,\left[D^{\alpha_{n}},\left[\mathcal{O}^{E}\left(\mathbf{k},t\right)\rho^{E}(\mathbf{k},t)\right]...\right]\right],
\]
The two terms in the expression (\ref{eq:SUM_RULE_TERM}) amount to
\[
\sum_{s}\int\frac{d^{d}\mathbf{k}}{\left(2\pi\right)^{d}}\nabla_{\mathbf{k}}\mathrm{G}_{ss}(\mathbf{k},t)-i\int\frac{d^{d}\mathbf{k}}{\left(2\pi\right)^{d}}\mathrm{Tr\left[\xi^{\alpha},G(\mathbf{k},t)\right]=0.}
\]
The first term integrates to zero due to the periodicity of $\mathrm{G}(\mathbf{k},t)$
in reciprocal space. As for the second one, the trace of any commutator
of two matrices is zero, which follows from the cyclic invariance
of the trace. 

The cancellation of all terms in the sum on the RHS of Eq\@.\,(\ref{eq:AVERSA_SUM})
constitute the sum rules referred to by Aversa and Sipe. They are
an order by order formulation of the result that in an integral over
the FBZ of a function with the period of the reciprocal lattice, the
argument of the integrand can be shifted without changing the integral. 

Nevertheless, it is clear from this formulation that these extra terms
exist only if, starting from the velocity gauge, we try to reduce
our expressions to the ones in the length gauge.

\section{Current response in a centrosymmetric material\label{sec:Current-response-in}}

For a centrosymmetric crystal, the spatial inversion operator $\mathcal{P}$
commutes with the crystal Hamiltonian, $\mathcal{H}$, 
\begin{align}
\left[\mathcal{P},\mathcal{H}\right] & =0.\label{eq:PARITY}
\end{align}
and the solutions to the $\mathbf{k}$-dependent Hamiltonian, $u_{\mathbf{k}s}(-\mathbf{r})$
and $u_{-\mathbf{k}s}(\mathbf{r})$, are related by a phase phase
factor,
\begin{equation}
u_{\mathbf{k}s}(-\mathbf{r})=e^{i\mu_{\mathbf{k}s}}\,u_{-\mathbf{k}s}(\mathbf{r}).
\end{equation}
If we take $\mathbf{r}\rightarrow-\mathbf{r}$ in the integral that
defines the Berry connection, Eq.\,(\ref{eq:CH10}), we can determine
how it behaves when we exchange the sign of the crystal momentum,
\begin{align}
\boldsymbol{\xi}_{\mathbf{k}ss'} & =\frac{i}{v_{C}}\int_{uc}d^{d}\mathbf{r}\,u_{\mathbf{k}s}^{*}(-\mathbf{r})\nabla_{\mathbf{k}}u_{\mathbf{k}s'}(-\mathbf{r})\nonumber \\
 & =-e^{i(\mu_{\mathbf{k}s'}-\mu_{\mathbf{k}s})}\left[\boldsymbol{\xi}_{-\mathbf{k}ss'}+\nabla_{\mathbf{k}}\mu_{\mathbf{k}s'}\delta_{ss'}\right],
\end{align}
and,
\begin{align}
\boldsymbol{\xi}_{-\mathbf{k}ss'} & =-e^{i(\mu_{\mathbf{k}s}-\mu_{\mathbf{k}s\text{'}})}\left[\boldsymbol{\xi}_{\mathbf{k}ss'}+\nabla_{\mathbf{k}}\mu_{\mathbf{k}s'}\delta_{ss'}\right].
\end{align}
This determines the transformation law for the covariant derivative:
\begin{align}
\mathbf{D}_{-\mathbf{k}ss'} & =-e^{i(\mu_{\mathbf{k}s}-\mu_{\mathbf{k}s\text{'}})}\left[\mathbf{D}_{\mathbf{k}ss'}+i\delta_{ss'}\nabla_{\mathbf{k}}\mu_{\mathbf{k}s'}\right].\label{eq:TLCOV}
\end{align}
We can now determine how the $\mathbf{k}$-dependent factor of $\rho^{(1)}$,
\begin{align}
\Pi_{\mathbf{k}ss'}^{(1),\alpha_{1}}(\omega) & =d_{\mathbf{k}ss'}(\omega)\bigl[D^{\alpha_{1}},\rho^{(0)}\bigr]_{\mathbf{k}ss'},\label{eq:PI_ONE}
\end{align}
transforms upon $\mathbf{k}\rightarrow-\mathbf{k}$, by recalling
that $f_{\mathbf{k}s}$ and $d_{\mathbf{k}ss'}(\omega)$ are even,
since they depend only on the band energies, 
\begin{align}
\Pi_{-\mathbf{k}ss'}^{(1),\alpha_{1}}(\omega) & =d_{-\mathbf{k}ss'}(\omega)\bigl[D^{\alpha_{1}},\rho^{(0)}\bigr]_{-\mathbf{k}ss'}\nonumber \\
 & =d_{\mathbf{k}ss'}(\omega)\left(D_{-\mathbf{k}ss'}^{\alpha_{1}}\,f_{\mathbf{k}s'}-f_{\mathbf{k}s}\,D_{-\mathbf{k}ss'}^{\alpha_{1}}\right)\nonumber \\
 & =-e^{i(\mu_{\mathbf{k}s}-\mu_{\mathbf{k}s\text{'}})}\,d_{\mathbf{k}ss'}(\omega)\,\bigl[D^{\alpha_{1}},\rho^{(0)}\bigr]_{\mathbf{k}ss'}\nonumber \\
 & =-e^{i(\mu_{\mathbf{k}s}-\mu_{\mathbf{k}s\text{'}})}\Pi_{\mathbf{k}ss'}^{(1),\alpha_{1}}(\omega).\label{eq:TLCOM}
\end{align}
This can be extended to the higher order contributions of the RDM,
in particular to the second order one, easily extracted from Eq.\,(\ref{eq:RECURSIVE1}),
\begin{align}
\Pi_{\mathbf{k}ss'}^{(2),\alpha_{1}\alpha_{2}}(\omega,\omega_{1}) & =d_{\mathbf{k}ss'}(\omega)\bigl[D^{\alpha_{1}},\Pi^{(1),\alpha_{2}}(\omega-\omega_{1})\bigr]_{\mathbf{k}ss'}.
\end{align}
It follows from Eqs.\,(\ref{eq:TLCOV}) and (\ref{eq:TLCOM}) that
\begin{align}
\Pi_{-\mathbf{k}ss'}^{(2),\alpha_{1}\alpha_{2}}(\omega,\omega_{1}) & =e^{i(\mu_{\mathbf{k}s}-\mu_{\mathbf{k}s\text{'}})}\Pi_{\mathbf{k}ss'}^{(2),\alpha_{1}\alpha_{2}}(\omega,\omega_{1}).
\end{align}
This object picks up the same $\mathbf{k}$-space phase factors, but
unlike its first order counterpart, the sign does not change. Combining
this with the transformation law for the velocity matrix element,
\begin{align}
\mathbf{v}_{\mathbf{k}s's} & \rightarrow\mathbf{v}_{-\mathbf{k}s's}\nonumber \\
 & =-e^{i(\mu_{\mathbf{k}s'}-\mu_{\mathbf{k}s})}\mathbf{v}_{\mathbf{k}s's}.\label{eq:VELOK}
\end{align}
we see that the integrand in the FBZ integral, $v_{\mathbf{k}s's}^{\beta}\Pi_{\mathbf{k}ss'}^{(2),\alpha_{1}\alpha_{2}}(\omega,\omega_{1})$
is an odd function of $\mathbf{k}$, the second order current vanishes. 

This argument also carries for an arbitrary order $n$. The \textbf{$\mathbf{k}$}-parity
of $v_{\mathbf{k}s's}^{\beta}\Pi_{\mathbf{k}ss'}^{(n),\alpha_{1}(...)\alpha_{n}}$
is determined by its number of covariant derivatives in $\Pi_{\mathbf{k}ss'}^{(n),\alpha_{1}(...)\alpha_{n}}$.
For $n$ even, the integrand is odd under $\mathbf{k}\rightarrow-\mathbf{k}$,
so even order contributions to the current vanish in a centrosymmetric
material.

\section{Deriving the expressions: (\ref{eq:1})-(\ref{eq:0})}

Consider $\Pi_{1}$, the collection of terms with only one intra band
factor, where we have used the two-band character of the monolayer
graphene, Eq.\,(\ref{eq:TRACE}),\begin{widetext}

\begin{align}
\Pi_{1}^{(3),\beta\alpha_{1}\alpha_{2}\alpha_{3}} & =i\sum_{r\,s's}v_{\mathbf{k}s's}^{\beta}\biggl\{\frac{1}{\hbar\omega-\epsilon_{\mathbf{k}ss'}}\nabla_{\mathbf{k}}^{\alpha_{1}}\left(\frac{1}{\hbar(\omega-\omega_{1})-\epsilon_{\mathbf{k}ss'}}\left(\delta_{r\bar{s'}}\frac{\xi_{\mathbf{k}s\bar{s'}}^{\alpha_{2}}\xi_{\mathbf{k}\bar{s'}s'}^{\alpha_{3}}f_{\mathbf{k}s'\bar{s'}}}{\hbar(\omega-\omega_{[2]})-\epsilon_{\mathbf{k}\bar{s'}s'}}-\delta_{r\bar{s}}\frac{\xi_{\mathbf{k}s\bar{s}}^{\alpha_{3}}\xi_{\mathbf{k}\bar{s}s'}^{\alpha_{2}}f_{\mathbf{k}\bar{s}s}}{\hbar(\omega-\omega_{[2]})-\epsilon_{\mathbf{k}s\bar{s}}}\right)\right)\nonumber \\
 & +\frac{1}{\hbar\omega-\epsilon_{\mathbf{k}ss'}}\left(\delta_{r\bar{s'}}\frac{\xi_{\mathbf{k}s\bar{s'}}^{\alpha_{1}}}{\hbar(\omega-\omega_{1})-\epsilon_{\mathbf{k}\bar{s'}s'}}\nabla_{\mathbf{k}}^{\alpha_{2}}\left(\frac{\xi_{\mathbf{k}\bar{s'}s'}^{\alpha_{3}}f_{\mathbf{k}s'\bar{s'}}}{\hbar(\omega-\omega_{[2]})-\epsilon_{\mathbf{k}\bar{s'}s'}}\right)-\delta_{r\bar{s}}\frac{\xi_{\mathbf{k}\bar{s}s'}^{\alpha_{1}}}{\hbar(\omega-\omega_{1})-\epsilon_{\mathbf{k}s\bar{s}}}\nabla_{\mathbf{k}}^{\alpha_{2}}\left(\frac{\xi_{\mathbf{k}s\bar{s}}^{\alpha_{3}}f_{\bar{s}s}}{\hbar(\omega-\omega_{[2]})-\epsilon_{\mathbf{k}s\bar{s}}}\right)\right)\nonumber \\
 & +\frac{1}{\hbar\omega-\epsilon_{\mathbf{k}ss'}}\left(\delta_{r\bar{s'}}\frac{\xi_{\mathbf{k}s\bar{s'}}^{\alpha_{1}}\xi_{\mathbf{k}\bar{s'}s'}^{\alpha_{2}}}{\hbar(\omega-\omega_{1})-\epsilon_{\mathbf{k}\bar{s'}s'}}\frac{\nabla_{\mathbf{k}}^{\alpha_{3}}f_{\mathbf{k}s'\bar{s'}}}{\hbar(\omega-\omega_{[2]})}-\delta_{r\bar{s}}\frac{\xi_{\mathbf{k}s\bar{s}}^{\alpha_{2}}\xi_{\mathbf{k}\bar{s}s'}^{\alpha_{3}}}{\hbar(\omega-\omega_{1})-\epsilon_{\mathbf{k}s\bar{s}}}\frac{\nabla_{\mathbf{k}}^{\alpha_{3}}f_{\mathbf{k}\bar{s}s}}{\hbar(\omega-\omega_{[2]})}\right)\biggr\}.\label{eq:F2}
\end{align}
\end{widetext} For $s'=\bar{s}$, the two terms in each line cancel
out by application of the Berry connection properties, Eq.\,(\ref{eq:BC1})
and (\ref{eq:BC2}). This fixes $s'=s$, and the $\Pi_{1}$ contribution
reads as\begin{widetext}

\begin{align}
\Pi_{1}^{(3),\beta\alpha_{1}\alpha_{2}\alpha_{3}} & =\frac{i}{\hbar\omega}\sum_{s}v_{\mathbf{k}ss}^{\beta}\Bigl\{\frac{1}{\hbar(\omega-\omega_{1})}\nabla_{\mathbf{k}}^{\alpha_{1}}\left(\xi_{\mathbf{k}s\bar{s}}^{\alpha_{2}}\xi_{\mathbf{k}\bar{s}s}^{\alpha_{3}}\,f_{\mathbf{k}s\bar{s}}\left(\frac{1}{\hbar(\omega-\omega_{[2]})-\epsilon_{\mathbf{k}\bar{s}s}}+\frac{1}{\hbar(\omega-\omega_{[2]})-\epsilon_{\mathbf{k}s\bar{s}}}\right)\right)\nonumber \\
 & +\frac{\xi_{\mathbf{k}s\bar{s}}^{\alpha_{1}}}{\hbar(\omega-\omega_{1})-\epsilon_{\mathbf{k}\bar{s}s}}\nabla_{\mathbf{k}}^{\alpha_{2}}\left(\xi_{\mathbf{k}\bar{s}s}^{\alpha_{3}}f_{\mathbf{k}s\bar{s}}\left(\frac{1}{\hbar(\omega-\omega_{[2]})-\epsilon_{\mathbf{k}\bar{s}s}}+\frac{1}{\hbar(\omega-\omega_{[2]})-\epsilon_{\mathbf{k}s\bar{s}}}\right)\right)\\
 & +\frac{1}{\hbar(\omega-\omega_{[2]})}\xi_{\mathbf{k}s\bar{s}}^{\alpha_{1}}\xi_{\mathbf{k}\bar{s}s}^{\alpha_{2}}\left(\frac{1}{\hbar(\omega-\omega_{1})-\epsilon_{\mathbf{k}\bar{s}s}}+\frac{1}{\hbar(\omega-\omega_{1})-\epsilon_{\mathbf{k}s\bar{s}}}\right)\nabla_{\mathbf{k}}^{\alpha_{3}}f_{\mathbf{k}s\bar{s}}\Bigr\}.\label{eq:F4}
\end{align}
\end{widetext} The $\Pi_{0}$ portion of the current for the two
band material reduces to,\begin{widetext}
\begin{align}
\Pi_{0}^{(3),\beta\alpha_{1}\alpha_{2}\alpha_{3}} & =\sum_{r'r\,s's}v_{\mathbf{k}s's}^{\beta}\frac{1}{\hbar\omega-\epsilon_{\mathbf{k}ss'}}\biggl\{\frac{1}{\hbar\left(\omega-\omega_{1}\right)-\epsilon_{\mathbf{k}rs'}}\left(\delta_{r'\bar{s'}}\frac{\xi_{\mathbf{k}sr}^{\alpha_{1}}\xi_{\mathbf{k}r\bar{s'}}^{\alpha_{2}}\xi_{\mathbf{k}\bar{s'}s'}^{\alpha_{3}}f_{\mathbf{k}s'\bar{s'}}}{\hbar\left(\omega-\omega_{[2]}\right)-\epsilon_{\mathbf{k}\bar{s'}s'}}-\delta_{r'\bar{r}}\frac{\xi_{\mathbf{k}sr}^{\alpha_{1}}\xi_{\mathbf{k}r\bar{r}}^{\alpha_{3}}\xi_{\mathbf{k}\bar{r}s'}^{\alpha_{2}}f_{\mathbf{k}\bar{r}r}}{\hbar\left(\omega-\omega_{[2]}\right)-\epsilon_{\mathbf{k}r\bar{r}}}\right)\nonumber \\
 & -\frac{1}{\hbar\left(\omega-\omega_{1}\right)-\epsilon_{\mathbf{k}sr}}\left(\delta_{r'\bar{r}}\frac{\xi_{\mathbf{k}s\bar{r}}^{\alpha_{2}}\xi_{\mathbf{k}\bar{r}r}^{\alpha_{3}}\xi_{rs'}^{\alpha_{1}}f_{\mathbf{k}r\bar{r}}}{\hbar\left(\omega-\omega_{[2]}\right)-\epsilon_{\mathbf{k}\bar{r}r}}-\delta_{r'\bar{s}}\frac{\xi_{\mathbf{k}s\bar{s}}^{\alpha_{3}}\xi_{\mathbf{k}\bar{s}r}^{\alpha_{2}}\xi_{rs'}^{\alpha_{1}}f_{\mathbf{k}\bar{s}s}}{\hbar\left(\omega-\omega_{[2]}\right)-\epsilon_{\mathbf{k}s\bar{s}}}\right)\Bigr\}.\label{eq:F5}
\end{align}
\end{widetext} Since the Berry connection is even under $\mathbf{k}\rightarrow-\mathbf{k}$,
the velocity matrix element in band space, Eq.\,(\ref{eq:VELCOM1}),
is written as the sum of two contributions of opposite parity: an
odd intra band term and an even inter band term. The integration over
the FBZ carries cancels all odd terms, and so the intra band part
of $\mathbf{v}_{\mathbf{k}ss'}$ can be ignored: this fixes $s'=\bar{s}$,
and \begin{widetext}
\begin{align}
\Pi_{0}^{(3),\beta\alpha_{1}\alpha_{2}\alpha_{3}} & =\sum_{r\,s}v_{\mathbf{k}\bar{s}s}^{\beta}\frac{1}{\hbar\omega-\epsilon_{\mathbf{k}s\bar{s}}}\biggl\{\frac{1}{\hbar\left(\omega-\omega_{1}\right)-\epsilon_{\mathbf{k}r\bar{s}}}\left(\frac{\xi_{\mathbf{k}sr}^{\alpha_{1}}\xi_{\mathbf{k}rs}^{\alpha_{2}}\xi_{\mathbf{k}s\bar{s}}^{\alpha_{3}}f_{\mathbf{k}\bar{s}s}}{\hbar\left(\omega-\omega_{[2]}\right)-\epsilon_{\mathbf{k}s\bar{s}}}-\frac{\xi_{\mathbf{k}sr}^{\alpha_{1}}\xi_{\mathbf{k}r\bar{r}}^{\alpha_{3}}\xi_{\mathbf{k}\bar{r}\bar{s}}^{\alpha_{2}}f_{\mathbf{k}\bar{r}r}}{\hbar\left(\omega-\omega_{[2]}\right)-\epsilon_{\mathbf{k}r\bar{r}}}\right)\nonumber \\
 & -\frac{1}{\hbar\left(\omega-\omega_{1}\right)-\epsilon_{\mathbf{k}sr}}\left(\frac{\xi_{\mathbf{k}s\bar{r}}^{\alpha_{2}}\xi_{\mathbf{k}\bar{r}r}^{\alpha_{3}}\xi_{\mathbf{k}r\bar{s}}^{\alpha_{1}}f_{\mathbf{k}r\bar{r}}}{\hbar\omega_{1}-\epsilon_{\mathbf{k}\bar{r}r}}-\frac{\xi_{\mathbf{k}s\bar{s}}^{\alpha_{3}}\xi_{\mathbf{k}\bar{s}r}^{\alpha_{2}}\xi_{r\bar{s}}^{\alpha_{1}}f_{\mathbf{k}\bar{s}s}}{\hbar\left(\omega-\omega_{[2]}\right)-\epsilon_{\mathbf{k}s\bar{s}}}\right)\biggr\}.\label{eq:F6}
\end{align}
\end{widetext} Setting $r=s$ and using (\ref{eq:BC1}) and (\ref{eq:BC2}),
the first two terms of this expression cancel out; the last two cancel
when $r=\bar{s}$. 

Finally, $\Pi_{0}$ reduces to, \begin{widetext}
\begin{align}
\Pi_{0}^{(3),\beta\alpha_{1}\alpha_{2}\alpha_{3}} & =\frac{2}{\hbar\left(\omega-\omega_{1}\right)}\sum_{s}\frac{1}{\hbar\omega-\epsilon_{\mathbf{k}s\bar{s}}}v_{\mathbf{k}\bar{s}s}^{\beta}\xi_{\mathbf{k}s\bar{s}}^{\alpha_{1}}\xi_{\mathbf{k}\bar{s}s}^{\alpha_{2}}\xi_{\mathbf{k}s\bar{s}}^{\alpha_{3}}f_{\mathbf{k}\bar{s}s}\left\{ \frac{1}{\hbar\left(\omega-\omega_{[2]}\right)-\epsilon_{\mathbf{k}s\bar{s}}}+\frac{1}{\hbar\left(\omega-\omega_{[2]}\right)-\epsilon_{\mathbf{k}\bar{s}s}}\right\} .\label{eq:F7}
\end{align}
\end{widetext}

%\bibliographystyle{apsrev4-1}
%\bibliography{NLO}

%

\end{document}